\documentclass[a4paper,10pt]{article}
\usepackage{jheppub}

\usepackage[utf8]{inputenc}
\usepackage{amsmath,amssymb,bm,slashed,braket}
\usepackage{dsfont}
\usepackage{amsfonts}
\usepackage{bm,bbm}
\usepackage{graphicx}
\usepackage{slashed} 
\usepackage[dvipsnames,table]{xcolor}
\usepackage[normalem]{ulem}
\usepackage{soul}
\usepackage{siunitx} 
\usepackage{hyperref}
\hypersetup{colorlinks,citecolor= blue,linkcolor= blue, urlcolor=blue}
\usepackage{ulem}
\usepackage{array}
\usepackage{verbatim}
\usepackage{epsfig}
\usepackage{multirow, booktabs}
\usepackage{bbold}
\usepackage[version=4]{mhchem}
\usepackage[capitalise]{cleveref}
\usepackage{tabularray}
\usepackage{makecell}
\usepackage{orcidlink}

\newcommand{\calO}{\mathcal{O}}
\newcommand{\C}{ {\tt C} }
\newcommand{\tL}{ {\tt L} }
\newcommand{\tR}{ {\tt R} }

\newcommand{\tf}{\texorpdfstring}

\definecolor{maroon}{cmyk}{0, 0.87, 0.68, 0.32}
\definecolor{halfgray}{gray}{0.55}
\definecolor{ipython_frame}{RGB}{207, 207, 207}
\definecolor{ipython_bg}{RGB}{247, 247, 247}
\definecolor{ipython_red}{RGB}{186, 33, 33}
\definecolor{ipython_green}{RGB}{0, 128, 0}
\definecolor{ipython_cyan}{RGB}{64, 128, 128}
\definecolor{ipython_purple}{RGB}{170, 34, 255}

\usepackage{listings}
\lstdefinelanguage{iPython}{
    morekeywords={access,and,break,class,continue,def,del,elif,else,except,exec,finally,for,from,global,if,import,in,is,lambda,not,or,pass,print,raise,return,try,while},%
    morekeywords=[2]{abs,all,any,basestring,bin,bool,bytearray,callable,chr,classmethod,cmp,compile,complex,delattr,dict,dir,divmod,enumerate,eval,execfile,file,filter,float,format,frozenset,getattr,globals,hasattr,hash,help,hex,id,input,int,isinstance,issubclass,iter,len,list,locals,long,map,max,memoryview,min,next,object,oct,open,ord,pow,property,range,raw_input,reduce,reload,repr,reversed,round,set,setattr,slice,sorted,staticmethod,str,sum,super,tuple,type,unichr,unicode,vars,xrange,zip,apply,buffer,coerce,intern},%
    sensitive=true,%
    morecomment=[l]\#,%
    morestring=[b]',%
    morestring=[b]",%
    morestring=[s]{'''}{'''},% used for documentation text (mulitiline strings)
    morestring=[s]{"""}{"""},% added by Philipp Matthias Hahn
    morestring=[s]{r'}{'},% `raw' strings
    morestring=[s]{r"}{"},%
    morestring=[s]{r'''}{'''},%
    morestring=[s]{r"""}{"""},%
    morestring=[s]{u'}{'},% unicode strings
    morestring=[s]{u"}{"},%
    morestring=[s]{u'''}{'''},%
    morestring=[s]{u"""}{"""},%
    literate=
    {á}{{\'a}}1 {é}{{\'e}}1 {í}{{\'i}}1 {ó}{{\'o}}1 {ú}{{\'u}}1
    {Á}{{\'A}}1 {É}{{\'E}}1 {Í}{{\'I}}1 {Ó}{{\'O}}1 {Ú}{{\'U}}1
    {à}{{\`a}}1 {è}{{\`e}}1 {ì}{{\`i}}1 {ò}{{\`o}}1 {ù}{{\`u}}1
    {À}{{\`A}}1 {È}{{\'E}}1 {Ì}{{\`I}}1 {Ò}{{\`O}}1 {Ù}{{\`U}}1
    {ä}{{\"a}}1 {ë}{{\"e}}1 {ï}{{\"i}}1 {ö}{{\"o}}1 {ü}{{\"u}}1
    {Ä}{{\"A}}1 {Ë}{{\"E}}1 {Ï}{{\"I}}1 {Ö}{{\"O}}1 {Ü}{{\"U}}1
    {â}{{\^a}}1 {ê}{{\^e}}1 {î}{{\^i}}1 {ô}{{\^o}}1 {û}{{\^u}}1
    {Â}{{\^A}}1 {Ê}{{\^E}}1 {Î}{{\^I}}1 {Ô}{{\^O}}1 {Û}{{\^U}}1
    {œ}{{\oe}}1 {Œ}{{\OE}}1 {æ}{{\ae}}1 {Æ}{{\AE}}1 {ß}{{\ss}}1
    {ç}{{\c c}}1 {Ç}{{\c C}}1 {ø}{{\o}}1 {å}{{\r a}}1 {Å}{{\r A}}1
    {€}{{\EUR}}1 {£}{{\pounds}}1
    {^}{{{\color{ipython_purple}\^{}}}}1
    {=}{{{\color{ipython_purple}=}}}1
    {+}{{{\color{ipython_purple}+}}}1
    {*}{{{\color{ipython_purple}$^\ast$}}}1
    {/}{{{\color{ipython_purple}/}}}1
    {+=}{{{+=}}}1
    {-=}{{{-=}}}1
    {*=}{{{$^\ast$=}}}1
    {/=}{{{/=}}}1,
    literate=
    *{-}{{{\color{ipython_purple}-}}}1
     {?}{{{\color{ipython_purple}?}}}1,
    identifierstyle=\color{black}\ttfamily,
    commentstyle=\color{ipython_cyan}\ttfamily,
    stringstyle=\color{ipython_red}\ttfamily,
    keepspaces=true,
    showspaces=false,
    showstringspaces=false,
    rulecolor=\color{ipython_frame},
    frame=single,
    frameround={t}{t}{t}{t},
    framexleftmargin=6mm,
    numbers=left,
    numberstyle=\tiny\color{halfgray},
    backgroundcolor=\color{ipython_bg},
    basicstyle=\footnotesize\ttfamily,
    keywordstyle=\color{ipython_green}\ttfamily,
    aboveskip=1.2em,
    belowskip=1.2em,
}

\usepackage{adjustbox}
\newcommand{\befun}{ \dot{C}}
\definecolor{lightbg}{gray}{0.96}

\allowdisplaybreaks[4]
\def\lsim{\mathrel{\raise.3ex\hbox{$<$\kern-.75em\lower1ex\hbox{$\sim$}}}}
\def\gsim{\mathrel{\raise.3ex\hbox{$>$\kern-.75em\lower1ex\hbox{$\sim$}}}}

\usepackage[edges]{forest}
\usepackage[section]{placeins}
\definecolor{foldercolor}{RGB}{124,166,198}
\usepackage{listings}

\tikzset{pics/folder/.style={code={%
			\node[inner sep=0pt, minimum size=#1](-foldericon){};
			\node[folder style, inner sep=0pt, minimum width=0.3*#1, minimum height=0.6*#1, above right, xshift=0.05*#1] at (-foldericon.west){};
			\node[folder style, inner sep=0pt, minimum size=#1] at (-foldericon.center){};}
	},
	pics/folder/.default={20pt},
	folder style/.style={draw=foldercolor!80!black,top color=foldercolor!40,bottom color=foldercolor}
}

\forestset{is file/.style={edge path'/.expanded={%
			([xshift=\forestregister{folder indent}]!u.parent anchor) |- (.child anchor)},
		inner sep=1pt},
	this folder size/.style={edge path'/.expanded={%
			([xshift=\forestregister{folder indent}]!u.parent anchor) |- (.child anchor) pic[solid]{folder=#1}}, inner xsep=0.6*#1},
	folder tree indent/.style={before computing xy={l=#1}},
	folder icons/.style={folder, this folder size=#1, folder tree indent=3*#1},
	folder icons/.default={12pt},
}

\title{RGE solver for the complete dim-7 SMEFT interactions and its application to $0\nu\beta\beta$ decay}

\author[a,b]{Yi Liao\,\orcidlink{0000-0002-1009-5483},}
\emailAdd{liaoy@m.scnu.edu.cn}
\affiliation[a]{State Key Laboratory of Nuclear Physics and
Technology, Institute of Quantum Matter, South China Normal
University, Guangzhou 510006, China}
\affiliation[b]{Guangdong Basic Research Center of Excellence for
Structure and Fundamental Interactions of Matter, Guangdong
Provincial Key Laboratory of Nuclear Science, Guangzhou
510006, China}
\author[a,b]{Xiao-Dong Ma\,\orcidlink{0000-0001-7207-7793},}
\emailAdd{maxid@scnu.edu.cn}
\author[a,b]{Hao-Lin Wang\,\orcidlink{0000-0002-2803-5657},}
\emailAdd{whaolin@m.scnu.edu.cn}
\author[a,b,c]{and Xiang Zhao\,\orcidlink{0009-0008-6024-7722}}
\emailAdd{zhaox88@mail2.sysu.edu.cn}
\affiliation[c]{School of Physics and Astronomy, Sun Yat-sen University, Zhuhai 519082, China}

\abstract{
We present an automatic renormalization group equations (RGEs) solver, \texttt{D7RGESolver}, designed for the precise numerical solution of one-loop RGEs of dimension-7 (dim-7) operators within the standard model effective field theory (SMEFT). This tool is capable of calculating the RGE effects of dim-5 and dim-7 SMEFT operators between any two scales above the electroweak scale. We take the nuclear neutrinoless double beta ($0\nu\beta\beta$) decay process as an example to appreciate the importance of the running effects in phenomenological studies. 
Our analysis demonstrates that $0\nu\beta\beta$ decay can constrain nearly all dim-7 SMEFT operators involving first-generation leptons that violate lepton number by two units, after accounting for RGE effects. 
Specifically, we have placed meaningful constraints on 55 dim-7 Wilson coefficients, 
compared to only 10 from a simple tree-level analysis.   
Certain operators exhibit much stricter constraints when RGE effects are included, especially for the operators that mix with neutrino mass operators. We provide a complete code documentation for \texttt{D7RGESolver}, along with examples of its usage and interfacing with external automated codes for calculating $0\nu\beta\beta$ decay. The \texttt{D7RGESolver} code is available at: 
\href{https://github.com/ZhaoXiang210/D7RGESolver}{Github: D7RGESolver}}

\keywords{Standard Model Effective Field Theory, Renormalization Group Equations, Neutrinoless Double Beta Decay, \texttt{D7RGESolver} code}
\makeatletter
\gdef\@fpheader{}
\makeatother
\begin{document} 
	
\maketitle
\setcounter{page}{2}

%%%%%%%%%%%%%%%%%%%%%%%%
\section{Introduction}
\label{sec:intro}
%%%%%%%%%%%%%%%%%%%%%%%%
	
Despite the remarkable success of the standard model (SM) in describing elementary particles and their interactions, it fails to address several fundamental questions, such as the origin of tiny neutrino masses, the nature of dark matter, and matter-antimatter asymmetry of the universe. These unresolved issues strongly indicate the existence of physics beyond the SM (BSM). However, given our limited knowledge about the exact nature of new physics (NP), a more general and systematic approach—Effective Field Theory (EFT)—has emerged as a powerful framework to help us understand the NP in an indirect way.
	
EFT provides a bottom-up approach to studying NP beyond the SM at a lower energy scale, where NP effects are encapsulated by model-independent effective operators 
and their corresponding Wilson coefficients (WCs) organized in a series expansion. In the energy region between the electroweak scale and the NP scale, these operators are built by the SM fields and preserve the SM gauge symmetry $\rm SU(3)_c \times SU(2)_L \times U(1)_Y$, forming the standard model effective field theory (SMEFT). Due to the renormalization effect, the WCs of the SMEFT operators evolve as the energy scale changes, which is captured by the renormalization group equations (RGEs). In order to understand the heavy NP effects more precisely, it is thus important to take into account these RGE effects in low energy analyses by solving these coupled RGEs with a high precision.
	
For dimension-5 (dim-5) SMEFT interactions, there is only a single Weinberg operator~\cite{Weinberg:1979sa}, which contributes to Majorana neutrino mass after the electroweak symmetry breaking (EWSB). The one-loop RGE of the Weinberg operator has been given in~\cite{Antusch:2001ck}.
For dim-6 SMEFT operators, the complete and independent basis has been constructed in~\cite{Grzadkowski:2010es} and the corresponding RGEs at one-loop level were derived in Refs.~\cite{Jenkins:2013zja,Jenkins:2013wua,Alonso:2013hga}, which have been implemented into automatic RGE solvers in recent years, including \texttt{Dsixtools}~\cite{Celis:2017hod,Fuentes-Martin:2020zaz}, \texttt{wilson}~\cite{Aebischer:2018bkb}, and \texttt{RGESolver}~\cite{DiNoi:2022ejg}. Based on these tools, the RGE effects have been extensively incorporated into phenomenological analyses, for instance, see Refs.~\cite{Kumar:2018kmr,Datta:2019zca,Greljo:2022jac,Bigaran:2020jil,Bigaran:2019bqv} and references therein. 

As demonstrated in~\cite{Kobach:2016ami}, only the odd higher-dimension operators can violate lepton number by two units 
($\Delta L =2$). Since the lepton number violating (LNV) processes arising from dim-5 Weinberg operator are suppressed by the tiny masses of neutrinos, the next dim-7 operators become particularly intriguing. These operators not only generate neutrino masses at the dim-7 level but also give rise to a richer variety of LNV processes beyond those associated with the dim-5 operator.
The construction of the complete set of dim-7 SMEFT operators was carried out in~\cite{Lehman:2014jma,Liao:2016hru}, and the corresponding one-loop RGEs due to the SM interactions have been calculated originally by Refs.~\cite{Liao:2016hru, Liao:2019tep}, which were later explicitly written down in terms of an explicit flavor basis in \cite{Zhang:2023ndw}. 
Although automatic RGE solvers for dim-6 operators have developed for nearly a decade, an automated RGE solver for the full set of dim-7 SMEFT operators is still missing. In this work, we fill this gap and provide the tool, \texttt{D7RGESolver}, for the calculation of these RGEs. \texttt{D7RGESolver} is a {\it Python} tool designed to efficiently solve the complete RGEs for the WCs of dim-5 and dim-7 SMEFT operators.

Since the dim-5 and dim-7 SMEFT interactions are fundamentally linked to LNV processes, numerous studies have explored various LNV processes within this SMEFT framework, including neutrino masses~\cite{Babu:2001ex,Chala:2021juk,Fridell:2024pmw}, nuclear neutrinoless double beta ($0\nu\beta\beta$) decay~\cite{deGouvea:2007qla,Deppisch:2017ecm,Cirigliano:2017djv,Cirigliano:2018yza,Scholer:2023bnn,Fridell:2023rtr,Li:2023wfi},
the $\mu^-\to e^+$ conversion in nuclei \cite{Fridell:2023rtr,SINDRUMII:1998mwd,Berryman:2016slh},
LNV signals in collider searches~\cite{Fuks:2020zbm,Aoki:2020til,Fuks:2020att,Graesser:2022nkv,Fridell:2023rtr,Li:2023wfi}, and LNV meson and charged lepton decays~\cite{Liao:2019gex,Li:2019fhz,Liao:2020roy,Zhou:2021lnl,
Liao:2021qfj,Fridell:2023rtr}. Besides, the transition magnetic moments of Majorana neutrinos induced by certain dim-7 operators have been searched for from neutrino-electron scattering experiments~\cite{Bell:2006wi,Giunti:2014ixa,Canas:2015yoa,AristizabalSierra:2021fuc,Magill:2018jla} as well as astrophysical observations~\cite{Raffelt:1998xu}.

In those previous works, only a subset of the dim-7 RGEs has been incorporated into the phenomenological analysis to account for the running effects~\cite{Liao:2019gex,Liao:2019tep,Scholer:2023bnn}. Given the significance of the complete RGEs, especially from the SM Yukawa sector due to the large top Yukawa coupling, it is crucial to include the full RGE effects in the examination of all relevant processes. 
Thus, developing an automatic tool,
\texttt{D7RGESolver}, for accurately solving the RGEs of those SMEFT operators is an important step to track the evolution of these WCs and understand their phenomenological consequences. Since the nuclear $0\nu\beta\beta$ decay process is the most critical for testing the Majorana nature of neutrinos and has been experimentally searched for vigorously \cite{KamLAND-Zen:2022tow,KamLAND-Zen:2024eml,Gomez-Cadenas:2019sfa,LEGEND:2017cdu,nEXO:2017nam,Han:2017fol,Armengaud:2019loe,Paton:2019kgy},  we will take this process as an example to illustrate the importance of the complete RGEs solved by \texttt{D7RGESolver}.

In previous studies of $0\nu\beta\beta$ decay, only a small subset of RGEs for the dim-7 $\Delta L=2$ interactions has been taken into account \cite{Liao:2019tep}, which was later incorporated into the automatic code--\texttt{$\nu$DoBe} \cite{Scholer:2023bnn}.
\texttt{$\nu$DoBe} can be used to calculate the half-life of $0\nu\beta\beta$ decay once the SMEFT operators and their WCs are given. However, our analysis from solving the complete RGEs reveals that the contributions neglected in \cite{Scholer:2023bnn} are phenomenologically significant. In particular, two aspects are not included in the \texttt{$\nu$DoBe} package.
On the one hand, the mixing between the dim-5 and dim-7 neutrino mass operators and other dim-7 operators is omitted. On the other hand, corrections in the RGEs from Yukawa interactions are disregarded, leading to the absence of the mixing among operators involving different generations. In this work, by applying  \texttt{D7RGESolver} to $0\nu\beta\beta$ decay, we demonstrate that a larger set of dim-7 LNV operators with various generation structures can be constrained by $0\nu\beta\beta$ decay due to the mixing induced from the full RGEs. The constraints on the relevant WCs vary significantly, depending on the specific operator structure and the flavors of the quark fields involved.
	
The remaining parts of the paper are organized as follows: In the next section, we introduce the dim-7 SMEFT operators and their RGEs, along with the notations adopted for the numerical solver \texttt{D7RGESolver}. In \cref{sec:RGEdim7}, we illustrate how to install and use \texttt{D7RGESolver} with some examples. In \cref{Sec:apply}, we apply \texttt{D7RGESolver} to $0\nu\beta\beta$ decay and interface it with \texttt{$\nu$DoBe} to compute $0\nu\beta\beta$ decay that is followed by numerical constraints on the relevant WCs due to the complete running effects.  Finally, in \cref{sec:summary}, we summarize our results. 
\cref{sec:full_RGEs} collects the RGEs of the SM parameters and the WCs of dim-5 and dim-7 SMEFT operators.

%%%%%%%%%%%%%%%%%%%%%%%%%%%%%%%%%%%%%%%%%%%%%%%%%%%
\section{Dim-7 SMEFT operator basis and RGEs}
\label{sec:smeft & RGEs}
%%%%%%%%%%%%%%%%%%%%%%%%%%%%%%%%%%%%%%%%%%%%%%%%%%%

The SMEFT works between the electroweak scale, $\Lambda_{\tt EW}$, and some unknown NP scale ($\Lambda$) which is much higher than $\Lambda_{\tt EW}$. It extends the SM Lagrangian with a tower of higher-dimensional local operators and their corresponding unknown WCs, which is organized according to the canonical dimensions of the operators. Those higher-dimensional operators are built out of the SM fields that satisfy the SM gauge symmetries $\rm SU(3)_c \times SU(2)_L \times U(1)_Y$. 
Generically, the SMEFT Lagrangian takes the form 
\begin{align}
\label{eq:LSMEFT}
{\cal L}_{\tt SMEFT}  = {\cal L}_{\tt SM} 
+ \sum_{i} C_{5}^i \calO_i^{(5)}
+ \sum_{i} C_{6}^i \calO_i^{(6)}
+ \sum_{i} C_{7}^i \calO_i^{(7)}
+\cdots, 
\end{align} 
where ${\cal L}_{\tt SM}$ stands for the SM Lagrangian and $\calO_i^{(d)}$ ($d\geq 5$) denotes a dim-$d$ SMEFT operator with $i$ indexing operators at this dimension. Its associated WC, $C_d^i$, encapsulates dynamics of heavy NP and carries an inverse mass dimension of $d-4$. To facilitate analysis, $C_d^i$ is typically  expressed in terms of a dimensionless coupling $c_d^i$ and an unknown NP scale $\Lambda$, as $C_d^i \equiv c_d^i/\Lambda^{d-4}$.  

For later convenience, we start with our conventions for the SM Lagrangian. The left-handed lepton and quark doublets are denoted by $L$ and $Q$, respectively, while the right-handed up-type quarks, down-type quarks, and charged leptons are represented by $u$, $d$, and $e$. The Higgs doublet is labeled by $H$, and the gauge bosons for the SM groups are denoted by $G_\mu^A,~W_{\mu}^I,~B_\mu$. Then the SM Lagrangian takes the form, 
\begin{align}
\label{eq:LSM}
\nonumber
\mathcal{L}_{\tt SM} & = 
- \frac{1}{4} G^{A}_{\mu\nu} G^{A\mu\nu} 
- \frac{1}{4} W^I_{\mu\nu} W^{I\mu\nu} 
- \frac{1}{4} B_{\mu\nu} B^{\mu\nu} 
+ ( D_\mu H)^\dagger (D^\mu H) 
- \mu_h^2 H^\dagger H 
- \lambda ( H^\dagger H)^2 
\\
& + \sum_{\psi=Q,L,u,d,e} \overline{\psi} i \slashed{D} \psi
- \big( 
 [Y_{u}]_{pr} \overline{Q_{p}} \tilde{H} u_{r} 
+ [Y_{d}]_{pr} \overline{Q_{p}} H d_{r}
+ [Y_l]_{pr} \overline{L_{p}} H e_{r } 
+ {\rm h.c.} 
\big)\;,
\end{align}
where the covariant derivative is defined by 
$D_\mu \equiv \partial_\mu - i g^{\prime} Y B_\mu 
- i g T^I W^I_\mu - i g_s T^A G^A_\mu$ with $g^{\prime},~g,~g_s$ representing the gauge couplings for the three gauge groups. $Y_{u,d,l}$ denote the associated Yukawa coupling matrices in three-generation space,  and $p,r,s,t=1,2,3$ are used to label the generations.

In this work, we focus on the dim-5 and dim-7 operators which provide the dominant LNV interactions related to $\Delta L=2$ processes. At dim 5, there is only a single operator \cite{Weinberg:1979sa}, which we denote as,
\begin{align}
\label{eq:dim-5}
\calO_{LH5}^{pr} = \epsilon_{ij} \epsilon_{mn} (\overline{L^{\C,i}_p} L^m_r) H^j H^n.
\end{align}
For the dim-7 operators, the first systematic study was conducted in \cite{Lehman:2014jma}, and the complete and independent operator basis was later provided in \cite{Liao:2016hru} by removing redundant operators from the previous work. Without enumerating fermion generations, there are $12$ operators with $\Delta L=\pm 2$ and 6 operators that violate both baryon and lepton numbers by one unit while keeping their sum conserved, i.e., $\Delta B=- \Delta L=\pm 1$. These operators are summarized in the second column of \cref{tab:dim7_SMEFT_OP}.
Since one always copes with specific fermions in physics applications, it was realized in \cite{Liao:2019tep} and then in \cite{Zhang:2023ndw} that a flavor-specific basis of operators with manifest flavor symmetries should better be employed. In this work
we adopt the operators presented in \cite{Liao:2019tep, Zhang:2023ndw} to organize the RGEs, and implement them into the numerical code \texttt{D7RGESolver}. The naming conventions and their interrelations between the two bases are also shown in the table.  
It should be noted that $\calO_{\bar  e LLLH}^{prst}
= \calO_{\bar  e LLLH}^{{\tt(S)},prst}+\calO_{\bar  e LLLH}^{{\tt(A)},prst}+\calO_{\bar  e LLLH}^{{\tt(M)},prst}$, which follows from the relation $(\calO_{\bar  e LLLH}^{prst}+r\leftrightarrow t)-r\leftrightarrow s=0$ \cite{Liao:2019tep}.

%%%%%%%%%%%%%%%
\begin{table}[t] 
\Large
\center
\resizebox{\linewidth}{!}{
\renewcommand\arraystretch{1.2}
\begin{tabular}{| c | l | l | l |c|}
\hline
\multicolumn{5}{|c|}{\cellcolor{gray!20}  
Dim-7 SMEFT operators: $(\Delta B,\Delta L)=(0,2)$} 
\\
\hline
Classes 
&\multicolumn{1}{|c|}{Original operator basis in \cite{Lehman:2014jma,Liao:2016hru}}
& {\makecell[c]{Basis in \\ \cite{Liao:2019tep,Zhang:2023ndw}}}
&\multicolumn{1}{|c|}{Relations of the two different notations}   
&{\makecell[c]{WC name in \\ \texttt{D7RGESolver}}}
\\
\hline
$\psi^2H^4$ 
& $\calO^{pr}_{L H}= \epsilon_{ij} \epsilon_{mn}  
\big(\overline{L^{i \C}_{p}} L^m_{r} \big) H^j H^n (H^\dagger H)$  
& $\calO^{pr}_{L H}$ 
&same
& {\tt LH\_pr} 
\\%
\hline
$\psi^2H^3D$ 
& $\calO^{pr}_{LeHD}=\epsilon_{ij} \epsilon_{mn} 
\big( \overline{L^{i \C}_{p}} \gamma_\mu e_{r}\big) 
H^j H^m i D^\mu H^n $ 
& $\calO^{pr}_{LeDH} $ 
& $=\calO^{pr}_{LeHD} $  
& {\tt LeDH\_pr}
\\%
\hline
\multirow{2}{*}{$\psi^2H^2 D^2$} 
& $\calO^{pr}_{LHD1} = \epsilon_{ij} \epsilon_{mn} \big(\overline{L^{i \C}_{p}}  D^\mu L^j_{r} \big) H^m D_\mu H^n$ 
& $ \calO^{pr}_{DLDH1} $ 
& $= \frac{1}{2}\big(\calO^{pr}_{LHD1}
+ \calO^{rp}_{LHD1} \big) $  
& {\tt DLDH1\_pr}
\\
& $\calO^{pr}_{LHD2} = \epsilon_{im} \epsilon_{jn} \big(\overline{L^{i \C}_{p} } D^\mu L^j_{r} \big) H^m D_\mu H^n$ 
& $ \calO^{pr}_{DLDH2} $ 
& $=\frac{1}{2} \big( \calO^{pr}_{LHD2} 
+ \calO^{rp}_{LHD2} \big)$ 
& {\tt DLDH2\_pr}
\\
\hline
\multirow{2}{*}{$\psi^2H^2 X$}
& $ \calO^{pr}_{L HB}=\epsilon_{ij} \epsilon_{mn} 
\big( \overline{L^{i \C}_{p}} \sigma_{\mu\nu} L^m_{r} \big) H^j H^n B^{\mu\nu}$  
& $\calO^{pr}_{L HB}$ 
& same   
& {\tt LHB\_pr}
\\%
& $\calO^{pr}_{LHW}=\epsilon_{ij} (\epsilon\tau^I)_{mn}
\big(\overline{L^{i \C}_{p}} \sigma_{\mu\nu} L^m_{r} \big) H^j H^n W^{I \mu\nu}$&$\calO^{pr}_{L HW} $ 
& same  
& {\tt LHW\_pr}
\\%
\hline
\multirow{7}{*}{$\psi^4H$}
&$\calO^{prst}_{\overline{e}LLL H} = 
\epsilon_{ij} \epsilon_{mn} 
\big( \overline{e_{p}} L^i_{r}\big) 
\big(\overline{L^{j \C}_{s}}L^m_{t} \big) H^n$ 
& $ \calO^{{\tt (S),}prst}_{\overline{e}LLL H} $ 
& $ =\frac{1}{6} \big( \calO^{prst}_{\overline{e}LLL H} + \calO^{pstr}_{\overline{e}LLL H} + \calO^{ptrs}_{\overline{e}LLL H} 
\big) + s\leftrightarrow t$ 
& {\tt eLLLHS\_prst}
\\
& 
& $\calO^{{\tt (A),}prst}_{\overline{e}LLL H} $ 
& $=\frac{1}{6} \big( \calO^{prst}_{\overline{e}LLL H} + \calO^{pstr}_{\overline{e}LLL H}+ \calO^{ptrs}_{\overline{e}LLL H}
\big)- s\leftrightarrow t$  
& {\tt eLLLHA\_prst}
\\
& 
& $\calO^{{\tt (M),} prst}_{\overline{e}LLL H} $ 
& $=\frac{1}{3} \big( \calO^{prst}_{\overline{e}LLL H} + \calO^{psrt}_{\overline{e}LLL H} \big)
-t\leftrightarrow r$   
& {\tt eLLLHM\_prst}
\\
& $\calO^{prst}_{\overline{d}L Q L H1}=
\epsilon_{ij} \epsilon_{mn} 
\big( \overline{d_{p}} L^i_{r} \big) 
\big( \overline{Q^{j\C}_{s}}L^m_{t}\big) H^n $ 
& $\calO^{prst}_{\overline{d}L Q L H1} $ 
& same  
& {\tt dLQLH1\_prst}
\\% 
& $\calO^{prst}_{\overline{d}L Q L H2}=
\epsilon_{im} \epsilon_{jn} 
\big( \overline{d_{p}} L^i_{r} \big) 
\big( \overline{Q^{j \C}_{s}}L^m_{t} \big) H^n$ 
& $\calO^{prst}_{\overline{d}L Q L H2} $  
& same
& {\tt dLQLH2\_prst}
\\% 
& $\calO^{prst}_{\overline{d}L ueH}=
\epsilon_{ij} \big( \overline{d_{p}} L^i_{r} \big) 
\big(\overline{u_{s}^\C} e_{t} \big) H^j$ 
& $\calO^{prst}_{\overline{d}L ueH} $ 
& same 
& {\tt dLueH\_prst}
\\%
& $\calO^{prst}_{\overline{Q} u LL H}=
\epsilon_{ij} \big(\overline{Q_{p}} u_{r} \big) 
\big(\overline{L^{\C}_{s}}  L^i_{t} \big) H^j$ 
& $\calO^{prst}_{\overline{Q} u LL H} $ 
& same  
& {\tt QuLLH\_prst}
\\% 
\hline
$\psi^4D$
& $\calO^{prst}_{\overline{d} uLL D} =
\epsilon_{ij}\big(\overline{d_{p}}\gamma_\mu u_{r}\big)
\big(\overline{L^{i \C}_{s}} i D^\mu L^j_{t}\big)$ 
& $\calO^{prst}_{\overline{d}uLDL}$ 
& $ =\frac{1}{2} \big( \calO^{prst}_{\overline{d} uLL D} + \calO^{prts}_{\overline{d} uLL D} \big) $
& {\tt duLDL\_prst}
\\%%%%% 
\hline
\multicolumn{5}{|c|}{\cellcolor{gray!20} 
Dim-7 SMEFT operators: $(\Delta B,\Delta L)=(1,-1)$ }
\\% 
\hline            
\multirow{4}{*}{$\psi^4 H$}
& $\calO^{prst}_{\overline{L}dud\widetilde{H}} =
\epsilon_{\alpha\beta\gamma}
\big(\overline{L_{p}} d^{\alpha}_{r} \big) 
\big(\overline{u^{\beta \C}_{s}} d^{\gamma}_{t}\big) \widetilde{H} $
&$\calO^{prst}_{\overline{L}dud\widetilde{H}}$
&same
& {\tt LdudH\_prst}
\\
& $\calO^{prst}_{\overline{L} dddH} =
\epsilon_{\alpha\beta\gamma}
\big(\overline{L_{p}} d^{\alpha}_{r}\big) 
\big(\overline{d^{\beta\C}_{s}}d^{\gamma}_{t}\big)H$
& $\calO^{{\tt (M)},prst}_{\overline{L}dddH}$
& $ =\frac{1}{3} 
\big( \calO^{prst}_{\overline{L}dddH} 
+ \calO^{psrt}_{\overline{L}dddH} \big)
-s\leftrightarrow t$  
& {\tt LdddHM\_prst}
\\%
& 
$\calO^{prst}_{\overline{e}Qdd\widetilde{H}}  = \epsilon_{ij} \epsilon_{\alpha\beta\gamma}
\big( \overline{ e_{p}} Q^{i\alpha}_{r}\big) 
\big( \overline{d^{\beta \C}_{s}} d^{\gamma}_{t}\big) \widetilde{H}^j$
& $\calO^{prst}_{\overline{e}Qdd\widetilde{H}}$ 
& same   
& {\tt eQddH\_prst}
\\%
& $\calO^{prst}_{\overline{L}dQQ\widetilde{H}}=
\epsilon_{ij}\epsilon_{\alpha\beta\gamma} 
\big(\overline{L_{p}} d^{\alpha}_{r} \big) 
\big(\overline{Q^{\beta \C}_{s}} Q^{i\gamma}_{t}\big) \widetilde{H}^j$  
& $ \calO^{prst}_{\overline{L}dQQ\widetilde{H}}$ 
& same
& {\tt LdQQH\_prst}
\\% 
\hline
\multirow{2}{*}{$\psi^4D$}
& $\calO^{prst}_{\overline{e}dddD} = 
\epsilon_{\alpha \beta\gamma}
\big(\overline{e_{p}} \gamma_\mu d^{\alpha}_{r}\big) \big(\overline{d^{\beta \C}_{s}}i D^\mu d^{\gamma}_{t} \big)$
&$\calO^{prst}_{\overline{e} ddDd}  $   
& $ =\frac{1}{6} 
\big( \calO^{prst}_{\overline{e}dddD} + \calO^{ptrs}_{\overline{e}dddD} +  \calO^{pstr}_{\overline{e}dddD} \big) + s \leftrightarrow t $   
& {\tt eddDd\_prst}
\\%
& $\calO^{prst}_{\overline{L} Qdd D}=   \epsilon_{\alpha\beta\gamma}
\big(\overline{L_{p}} \gamma_\mu Q^{\alpha}_{r} \big)
\big(\overline{d^{\beta \C}_{s}} i D^\mu d^{\gamma}_{t} \big)$
& $\calO^{prst}_{\overline{L}QdDd}$ 
& $ =\frac{1}{2} 
\big( \calO^{prst}_{\overline{L} Qdd D} + \calO^{prts}_{\overline{L} Qdd D} \big) $   
& {\tt LQdDd\_prst}
\\
\hline
\end{tabular}
}
\caption{Summary of the dim-7 SMEFT operators, which are categorized into two subsets characterized by $(\Delta B,\Delta L)=(0,2)$ and $(\Delta B,\Delta L)=(1,-1)$. The Hermitian conjugate of the above operators multiplied by their WCs should be included in physical applications.}
\label{tab:dim7_SMEFT_OP}
\end{table}
%%%%%%%%%%%%%%%	

The RGE running of effective operators induced by SM interactions plays a crucial role in precision phenomenological analyses, especially for the LNV dim-5 and dim-7 interactions, given their origins at much higher energy scales. The one-loop RGEs for the dim-5 Weinberg operator were given in~\cite{Antusch:2001ck}.
The complete one-loop RGEs for the six dim-7 BNV operators were first computed in \cite{Liao:2016hru}, while the full one-loop SM corrections for the twelve  $\Delta L=2$ operators were initially provided in \cite{Liao:2019tep}. More recently, Ref.\,\cite{Zhang:2023ndw} adopted the basis with manifest flavor symmetries given in \cref{tab:dim7_SMEFT_OP} to recalculate the RGEs, expressing them in an explicit form and correcting an error in the lepton Yukawa term of the RGEs for the operators $\calO_{LHB,LHW}$ due to an insertion of the operator $\calO_{LeHD}$ in the earlier work. To consistently account for the RGE effects, the mixing induced by renormalization among operators of different dimensions may be significant. 
Ref.~\cite{Zhang:2023kvw} has computed the RGE evolution contributions to the dim-5 operator arising from the insertion of dim-7 operators. As will be demonstrated later, these effects play an important role in evaluating the constraints on dim-7 operators; therefore, we have also incorporated them into our code. Higher-order effects arising from multiple insertions of dim-5, -6, -7 interactions are phenomenologically suppressed and thus neglected in our analysis.
For reader's convenience, the complete one-loop RGEs for the dim-5 and dim-7 operators discussed above, together with the one-loop RGEs of the SM parameters, are collected in \cref{sec:full_RGEs}.
These expressions are encoded in the file \texttt{beta\_function.py} within \texttt{D7RGESolver} package, 
which will be described in detail in the next section.

In this work, we focus on numerically solving the one-loop RGEs for dim-5 and dim-7 operators. To this end, we present the automatic RGE solver, \texttt{D7RGESolver}, specifically developed for the precise numerical solution of one-loop RGEs for the dim-5 and dim-7 SMEFT operators. This package is capable of computing the RGE effects of both dim-5 and dim-7 SMEFT operators between any two energy scales above the electroweak scale. It is expected to serve as a valuable tool for precise studies of low-energy LNV and BNV processes. To implement these RGEs into \texttt{D7RGESolver}, we adopt the naming conventions for the WCs used in the {\tt wilson} package for the dim-6 RGEs~\cite{Aebischer:2018bkb}. These conventions are listed in the last column of \cref{tab:dim7_SMEFT_OP} for each corresponding flavor-specific operator. Similarly, the WC of the dim-5 operator in \cref{eq:dim-5} is denoted as \texttt{LH5\_pr} in \texttt{D7RGESolver}.

%%%%%%%%%%%%%%%%%%%%
\begin{table}[t]
\label{tab:WCs_dic}
\centering
\resizebox{\textwidth}{!}{
\renewcommand{\arraystretch}{1}
\begin{tabular}{|c|c|c|c|c|c|c|c|c|c|c|}
\hline
Class & \texttt{2F} & \texttt{2FS}& \texttt{2FA} & \texttt{4F3S} & \texttt{4F3A} & \texttt{4F2S} & \texttt{4F2A} &  \texttt{4F} & \texttt{4F3M1} & \texttt{4F3M2} \\
\hline
\makecell[c]{\large \#}&9 &6&3&30&3&54&27&81&24&24
\\
\hline
\multirow{6}{*}{\rotatebox{90}{
\large WC in the code~} }
& {\tt LeDH\_pr}
& {\tt LH5\_pr}
& {\tt LHB\_pr}
& {\tt eLLLHS\_prst}
& {\tt eLLLHA\_prst}
& {\tt duLDL\_prst}
&\cellcolor{gray!20}{\tt eQddH\_prst}
& {\tt dLQLH1\_prst}
& {\tt eLLLHM\_prst}
&\cellcolor{gray!20}{\tt LdddHM\_prst}
\\
& {\tt LHW\_pr}
& {\tt LH\_pr} &
& \cellcolor{gray!20}{\tt eddDd\_prst} & &\cellcolor{gray!20}{\tt LQdDd\_prst}
& &{\tt dLQLH2\_prst} & &
\\
&
& {\tt DLDH1\_pr} & & & & & 
&{\tt dLueH\_prst} & &
\\
&
& {\tt DLDH2\_pr} & &  & &
& &{\tt QuLLH\_prst} & &
\\
& &  & &  & & & 
&\cellcolor{gray!20}{\tt LdudH\_prst} & &
\\
& &  & & & & & 
&\cellcolor{gray!20}{\tt LdQQH\_prst} & &
\\
\hline
 \large sym.
& \textbackslash
& $\tt pr=rp$
& $\tt pr=-rp$
& $\makecell{\tt prst=prts \\
~~~~~~~\tt =ptsr }$
& $\makecell{\tt prst=-prts \\
~~~~~~~ \tt =-ptsr }$
& $\tt prst=prts$
& $\tt prst=-prts$
& \textbackslash 
& $\makecell{\tt prst=-ptsr  \\
\tt =psrt+prts }$
& $\makecell{\tt prst=-prts \\
\tt =psrt+ptsr }$
\\
\hline
\multirow{10}{*}{\rotatebox{90}{\large Adopted flavors}}
& \{\texttt{p}, \texttt{r}\} 
& \real{\{1, 1\}}
& \real{\{1, 2\}} 
& \{\texttt{p}, 1, 1, 1\} 
& {\{\texttt{p}, 1, 2, 3\}}
& \real{\{\texttt{p}, \texttt{r}, 1, 1\}} 
& \{\texttt{p}, \texttt{r}, 1, 2\} 
& \{\texttt{p}, \texttt{r}, \texttt{s}, \texttt{t}\} 
& \{\texttt{p}, 1, 1, 2\} 
& \{\texttt{p}, 1, 1, 2\} 
\\ 
&  
& \{1, 2\} 
& \{1, 3\} 
& \{\texttt{p}, 1, 1, 2\} &  
& {\{\texttt{p}, \texttt{r}, 1, 2\}} 
&  \{\texttt{p}, \texttt{r}, 1, 3\} &  
& \{\texttt{p}, 1, 2, 2\} 
& \{\texttt{p}, 2, 1, 2\} 
\\ 
&  
& \{1, 3\} 
& \{2, 3\} 
& \{\texttt{p}, 1, 1, 3\} &  
& {\{\texttt{p}, \texttt{r}, 1, 3\}} 
& \{\texttt{p}, \texttt{r}, 2, 3\} &  
& \{\texttt{p}, 1, 3, 2\} 
& \{\texttt{p}, 3, 1, 2\} 
\\ 
&  
& \real{\{2, 2\}} & 
& \{\texttt{p}, 1, 2, 2\} &  
& {\{\texttt{p}, \texttt{r}, 2, 2\}} &  &  
& \{\texttt{p}, 1, 1, 3\} 
& \{\texttt{p}, 1, 1, 3\} 
\\ 
&  
& \{2, 3\} & 
& \{\texttt{p}, 1, 2, 3\} &  
& \{\texttt{p}, \texttt{r}, 2, 3\} &  &  
& \{\texttt{p}, 1, 2, 3\} 
& \{\texttt{p}, 2, 1, 3\} 
\\ 
&  
& \{3, 3\} & 
& \{\texttt{p}, 1, 3, 3\} &  
& \real{\{\texttt{p}, \texttt{r}, 3, 3\}} &  &  
& \{\texttt{p}, 1, 3, 3\} 
& \{\texttt{p}, 3, 1, 3\} 
\\ 
&  &  & 
&\{\texttt{p}, 2, 2, 2\} &  &  &  &  
&\{\texttt{p}, 2, 2, 3\} 
&\{\texttt{p}, 2, 2, 3\} 
\\ 
&  &  & 
&\{\texttt{p}, 2, 2, 3\} & &  &  &  
&\{\texttt{p}, 2, 3, 3\} 
&\{\texttt{p}, 3, 2, 3\}
\\ 
&  &  & 
&\{\texttt{p}, 2, 3, 3\} &  &  &  &  &  & 
\\ 
&  &  & 
&\{\texttt{p}, 3, 3, 3\} &  &  &  &  &  &  
\\ 
\hline
\end{tabular}
}  
\caption{Independent WCs dictionary of dim-7 SMEFT operators adopted by \texttt{D7RGESolver}. The fourth row shows the flavor symmetry properties of all dim-7 SMEFT operators. 
In the last row, we list the independent generation combinations adopted by \texttt{D7RGESolver}, where the unspecified generation indices $\texttt{p}, \texttt{r}, \texttt{s}, \texttt{t}$ 
can be either 1, 2, or 3. } 
\label{tab:dict}
\end{table}
%%%%%%%%%%%%%%%%%%%%

Due to flavor symmetry of these operators, it is necessary to fix the independent generation indices in each operator to obtain unique numerical solutions. For this purpose, we summarize the flavor symmetry properties of the adopted operators in \cref{tab:dict}, along with the number of independent operators with three generations (second row). The elements listed in the last row of \cref{tab:dict} form a complete basis for dim-7 flavor-specific operators and are inputted in \texttt{D7RGESolver}. Any other combinations of fermion generation indices are redundant and not implemented in \texttt{D7RGESolver}; their running results can be directly obtained from those of the selected combinations.

%%%%%%%%%%%%%%%%%%%%%%%%%%%%%%%%%%%%%%
\section{Usage of {\tt D7RGESolver}}
\label{sec:RGEdim7}
%%%%%%%%%%%%%%%%%%%%%%%%%%%%%%%%%%%%%%

In this section, we introduce the basic usage of \texttt{D7RGESolver}, including its installation and basic commands to solve the full RGEs of dim-7 SMEFT operators.

%%%%%%%%%%%%%%%%%%%%%%%%%%%%%%%%%%%%%%
\subsection{Installation of \texttt{D7RGESolver}}
%%%%%%%%%%%%%%%%%%%%%%%%%%%%%%%%%%%%%%

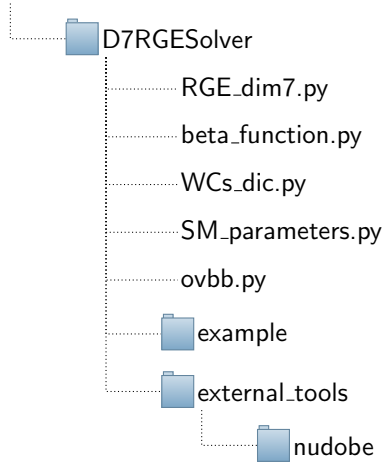
\begin{figure}[h!]
    \center
	\begin{forest}
	   for tree={font=\sffamily, grow'=0, folder indent=.9em, folder icons,
	   edge=densely dotted}
	   [
        [D7RGESolver,
	    [RGE\_dim7.py, is file]
		[beta\_function.py, is file]
		[WCs\_dic.py, is file]
		[SM\_parameters.py, is file]
		[ovbb.py, is file]
		[example]
		[external\_tools,[
		nudobe]]]
	   ]
	\end{forest}
	\caption{The structure of the \texttt{D7RGESolver} package.}
	\label{fig:directory_tree_RGE7}
\end{figure}

Users can download \texttt{D7RGESolver} package from \href{https://github.com/ZhaoXiang210/D7RGESolver}{Github: D7RGESolver} and put it in any directory with a valid {\it Python} environment. After installation, the package's file structure is displayed in \cref{fig:directory_tree_RGE7}.
			
Inside of the main directory, 
the \texttt{WCs\_dic.py} file stores all selected independent WCs of dim-5 and dim-7 SMEFT operators with explicit fermion generation indices as listed in \cref{tab:dict}. 
The \texttt{beta\_function.py} file contains the full set of beta functions for both the SM parameters and the chosen WCs of dim-5 and dim-7 operators, as compiled in \cref{sec:full_RGEs}. 
The \texttt{SM\_parameters.py} file stores the SM parameter inputs, which are summarized in \cref{tab:smpar}, and are consistent with the values used in the \texttt{wilson}~\cite{Aebischer:2018bkb} and \texttt{DsixTools 2.0}~\cite{Fuentes-Martin:2020zaz} packages.
The RGEs are numerically solved by using the functions defined in \texttt{RGE\_dim7.py}. 
Users can also flexibly include custom packages or scripts in the {\tt external\_tools} directory to interface with other tools, thus enabling direct import of the RGE-improved WCs into phenomenological analyses.
As an example, we provide an interface to the \texttt{$\nu$DoBe} package \cite{Scholer:2023bnn} (located in the \texttt{nudobe} folder), which automates the calculation of $0\nu\beta\beta$ decay within the EFT framework~\cite{Cirigliano:2017djv, Cirigliano:2018yza}.
The \texttt{ovbb.py} script is used to interface \texttt{D7RGESolver} with \texttt{nudobe}, enabling the computation of $0\nu\beta\beta$ decay and corresponding constraints on WCs. Several examples are provided in the \texttt{example} folder via \href{https://jupyter.org}{Jupyter notebook}, demonstrating how to use \texttt{D7RGESolver} and interface it with \texttt{nudobe}.

%%%%%%%%%%%%%%%%%%%%
\begin{table}[t]
\center
\resizebox{\textwidth}{!}{
\renewcommand{\arraystretch}{1.0}
\begin{tabular}{|ll|ll|ll|ll|ll|}
\hline 
\multicolumn{2}{|c|}{gauge couplings} & \multicolumn{2}{c|}{Quark masses} & \multicolumn{2}{c|}{Lepton masses} & \multicolumn{2}{c|}{EW parameters} & \multicolumn{2}{c|}{CKM parameters} 
\\
\hline
$g^{\prime}$ & 0.357562 
& $m_u$ & 0.00127\,GeV 
& $m_e$ & 0.000511\,GeV 
&  $\mu_h^2$ & $-8528.18\,\rm GeV^2$  
& $V_{u s}$ & 0.2243
\\
$g$ & 0.651451 
& $m_d$ & 0.00270\,GeV 
& $m_\mu$ & 0.1057\,GeV 
&  $\lambda$ & 0.140673  
& $V_{c b}$ & 0.04221 
\\
$g_s$ & 1.220293 
& $m_s$ & 0.0551\,GeV 
& $m_\tau$ & 1.777\,GeV &  &  
& $V_{u b}$ & 0.00362 
\\
&  & $m_c$ & 0.635\,GeV &  &  &  &  
& $\delta_{\text{CP}}$ &1.27  
\\
&  & $m_b$ & 2.85\,GeV &  &  & &  &  &  
\\
&  & $m_t$ & 169.0\,GeV  &  &  &  &  &  &  
\\
\hline
\end{tabular} }
\caption{The input values of the SM parameters adopted by \texttt{D7RGESolver} at $Z$-pole, $M_Z=91.1876\,\rm GeV$. }
\label{tab:smpar}
\end{table}
%%%%%%%%%%%%%%%%%%%%%

As the solution of these RGEs depends on the specific form of the Yukawa matrices, we consider two commonly used quark flavor bases in this work, namely, the up-quark flavor basis and the down-quark flavor basis \cite{Liao:2024xel}. 
In the up-quark flavor basis, both the left- and right-handed up-type quark fields and the right-handed down-type quark fields are already in their mass eigenstates. In this case, the weak eigenstates $d'_{\tL}$ and mass eigenstates $d_{\tL}$ of the left-handed down-type quarks are related by the Cabibbo-Kobayashi-Maskawa (CKM) matrix \cite{Cabibbo:1963yz, Kobayashi:1973fv} via $d'_{\tL}=V d_{\tL}$. 
Accordingly, the Yukawa matrix for up-type quarks is taken to be diagonal at the electroweak scale $\Lambda_{\tt EW}$ while the down-type Yukawa matrix incorporates the CKM matrix at $\Lambda_{\tt EW}$,
\begin{align}
\label{eq:up}
Y_u=\frac{\sqrt{2}}{v} {{M}_u},\quad
Y_d=\frac{\sqrt{2}}{v} V {{M}_d},
\end{align}
where ${M}_u = \text{diag}(m_u, m_c, m_t)$ and ${M}_d = \text{diag}(m_d, m_s, m_b)$ are respectively the diagonal mass matrices for up- and down-type quarks, and $v$ is the vacuum expectation value (VEV) of the Higgs field. In the down-quark flavor basis, the weak and physical mass eigenstates of left-handed up-type quarks are related by $u_\tL'=V^\dagger u_\tL$, and the quark Yukawa matrices at $\Lambda_{\tt EW}$ are
\begin{align}
\label{eq:down}
Y_u=\frac{\sqrt{2}}{v} V^{\dagger}{{M}_u},\quad Y_d=\frac{\sqrt{2}}{v}{{M}_d}.
\end{align}
In both quark flavor bases, the charged lepton Yukawa matrix $Y_l$ is taken to be diagonal at $\Lambda_{\tt EW}$
\begin{align}
Y_l=\frac{\sqrt{2}}{v} {{M}_e}\;,
\end{align}
where ${M}_e = \text{diag}(m_e, m_\mu, m_\tau)$ is the diagonal charged lepton mass matrix.
It is important to note that $Y_u$ ($Y_d$) in the up-quark (down-quark) flavor basis is defined to be diagonal only at $\Lambda_{\tt EW}$, but will generally become non-diagonal at other scales since their RGEs involve non-diagonal anomalous dimension matrices from the Yukawa interactions, see \cref{eq:Yurun} and \cref{eq:Ydrun}.
In contrast, the lepton Yukawa matrix $Y_l$ remains diagonal under the RGE running as clearly seen from \cref{eq:Ylrun}. 

%%%%%%%%%%%%%%%%%%%%%%%%%%%%%%%%%%%%%%%%%%%%%%%%%%%%%%
\subsection{Basic commands of \texttt{D7RGESolver}}
\label{sec:D7rgesolver}
%%%%%%%%%%%%%%%%%%%%%%%%%%%%%%%%%%%%%%%%%%%%%%%%%%%%%%

The \texttt{D7RGESolver} is capable of solving the RGEs in both directions, from lower to higher energy scales and vice versa. 
To solve the RGEs for dim-5 and dim-7 SMEFT operators using \texttt{D7RGESolver}, we highly recommend executing codes in a \href{https://jupyter.org}{Jupyter notebook} within the \texttt{D7RGESolver} directory.
As a demonstration example, we consider running WCs of the dim-5 operator $\calO_{LH5}^{11}$ and a dim-7 operator $\calO_{DLDH1}^{11}$ from a high scale (10 TeV) to the electroweak scale, $\Lambda_{\texttt{EW}}\equiv$ 80 GeV, a value used by $\nu \texttt{DoBe}$ and we adopt it for later numerical comparison. 
The process begins by importing relevant functions and specifying the input parameters. 
%%%%%%%%%%%%%%		
\begin{lstlisting}[language=iPython, numbers=none, 
				breaklines=true, 
				xleftmargin=2.5em, 
				xrightmargin=0.5em
				]
from RGE_dim7 import solve_rge, print_WCs
C_in = {"LH5_11": 1e-15+0j,"DLDH1_11": 1e-15+0j}       # Input the WCs
scale_in = 1e4                   # Input energy scale in units of GeV
scale_out = 80                   # Output energy scale in units of GeV
\end{lstlisting}
%%%%%%%%%%%%%%%
The inputs include values of the WCs in units of ${\rm GeV}^{d-4}$ (stored in the dictionary \texttt{C\_in}) at a given initial scale (\texttt{scale\_in}) and the value of the output energy scale (\texttt{scale\_out}). 
Both input and output scales are given in units of GeV and must exceed the electroweak scale $\Lambda_{\tt EW}$, which is set to $80~\rm GeV$ by default in \texttt{D7RGESolver}. 
\
The key numerical RGE solution is performed via the \texttt{solve\_rge} function, which supports two different integration methods. 
The default high-precision method (\texttt{method="integrate"}) employs \href{https://scipy.org}{SciPy}’s
\texttt{solve\_ivp} function to solve RGEs, and will be adopted for the following $0\nu\beta\beta$ study. Users may alternatively choose the leading-logarithmic approximation (\texttt{method="leadinglog"}) for rapid but less precise solutions. In addition, one can specify either the down- or up-quark flavor basis by using \texttt{basis="down"} or \texttt{basis="up"} to obtain the output results, with \texttt{basis="down"} being the default option.
In the down-quark flavor basis, the RGEs are solved by the following single-line code, 
%%%%%%%%%%%%%%%
\begin{lstlisting}[language=iPython, numbers=none, 
				breaklines=true, 
				xleftmargin=2.5em, 
				xrightmargin=0.5em
				]
C_out = solve_rge(scale_in, scale_out, C_in, basis="down", method="integrate") 
\end{lstlisting}
%%%%%%%%%%%%%%%%
Finally, users can print the evolved WCs in a more formatted form by using the function \texttt{prints\_WCs}:
\begin{lstlisting}[language=iPython, numbers=none, 
				breaklines=true, 
				xleftmargin=2.5em, 
				xrightmargin=0.5em
				]
print_WCs(C_out)
\end{lstlisting}
which will output results as follows:
{\small
\begin{verbatim}
## Wilson coefficients

**EFT:** `SMEFT`
| WC name | Value |
|--------------------|----------------------------------------------------|
| `LH5_11` | (5.51729127984111e-13-1.1697182471147425e-28j) |
| `LH_11` | (-2.7271477267102585e-18+1.9639321856872016e-33j) |
| `DLDH1_11` | (8.183376576714679e-16+1.539722194653356e-39j) |
| `DLDH2_11` | (8.253541710427263e-17-3.157387110573996e-39j) |
| `QuLLH_3311` | (-2.976790814412032e-17+6.616906663619594e-33j) |
| `LHW_11` | (-3.399190512252408e-18-1.1024467172929897e-33j) |
| ... | ... |
\end{verbatim}}
\normalsize
\noindent 
In the function \texttt{print\_WCs}, the WCs \texttt{LH5\_pr} and \texttt{LH\_pr} of dim-5 and dim-7 neutrino mass operators are always shown on top and the other non-zero WCs of dim-7 operators are arranged in the descending order of absolute values. 
Note that the WC $C_{LH5}^{11}$ at the output scale ($80\,\rm GeV$) is significantly larger than the input value at $\texttt{scale\_in}=10^4\,\rm GeV$. This enhancement arises from the RG mixing contribution due to the operator $\mathcal{O}_{DLDH1}^{11}$ which has a large input value for its WC $C_{DLDH1}^{11}$. According to  \cref{eq:RGE_Weinberg_dim5}, the correction is estimated to be $\delta {C_{LH5}^{11}}\sim 6g^2/(16\pi^2)\mu_h^2C_{DLDH1}^{11}\ln(80/10^4)$, which is very close to the above output value.

%%%%%%%%%%%%%%%%%%%%%%%%%%%%%%%%%%%%%%
\section{Application of \texttt{D7RGESolver} to \tf{$0\nu\beta\beta$}{0vbb} decay}
\label{Sec:apply}
%%%%%%%%%%%%%%%%%%%%%%%%%%%%%%%%%%%%%%

In this section, we take the pivotal $0\nu\beta\beta$ decay as an example to demonstrate the importance of a complete RGE evolution in  phenomenological studies. As will be seen later, the current $0\nu\beta\beta$ decay lifetime bound can constrain more WCs than a naive tree-level analysis, thanks to the significant operator mixing effects and sizable Yukawa corrections. 

%%%%%%%%%%%%%%%%%%%%%%%%%%%%%%%%%%%%%%
\subsection{EFT framework for \tf{$0\nu\beta\beta$}{0vbb} decay}
%%%%%%%%%%%%%%%%%%%%%%%%%%%%%%%%%%%%%%

We start from the low energy effective field theory (LEFT) description of $0\nu\beta\beta$ decay. In the LEFT framework, the relevant interactions are built by physical fields in mass eigenstate (except for the neutrino fields) that satisfy the unbroken $\rm SU(3)_c\times U(1)_{em}$ gauge symmetry of SM. For the $0\nu\beta\beta$ decay process, the relevant degrees of freedom include the up and down quarks, electron, and electron-type neutrino. The effective Lagrangian related to $0\nu\beta\beta$ decay takes the following general form,
\begin{align}
\label{eq:LEFT}
\mathcal{L}_{\tt LEFT}^{0\nu\beta\beta}
=\mathcal{L}^{(3)}_{\Delta L=2}
+\mathcal{L}^{(6)}_{\Delta L=2}
+\mathcal{L}^{(7)}_{\Delta L=2}
+\mathcal{L}^{(8)}_{\Delta L=2}
+\mathcal{L}^{(9)}_{\Delta L=2}+\cdots,
\end{align}
where the dim-3 and dim-6 through dim-9 LEFT interactions are responsible for the decay via the mass, long-distance, and short-distance mechanisms, respectively. 
Explicitly, the following interactions are matched from dim-5 and dim-7 SMEFT operators \cite{Cirigliano:2017djv},
\begin{subequations}
\label{eq:LEFT0vbb}
\begin{align}
\mathcal{L}^{(3)}_{\Delta L=2} & =-\frac{1}{2}
m_{\beta\beta}\,\overline{\nu_ {{\tt L},e}^\C}\nu_ {{\tt L},e}+{\rm h.c.}\;,
\\% 
\mathcal L^{(6)}_{\Delta L = 2} & = 
\sqrt{2} G_F
\Big[ C^{(6)}_{\textrm{VL}}   
(\overline u_\tL\gamma^\mu d_\tL) 
(\overline e_\tR\gamma_\mu \nu^{\C}_{\tL,e}) 
+ C^{(6)}_{\textrm{VR}}
(\overline u_\tR \gamma^\mu d_\tR)
(\overline e_\tR\gamma_\mu\nu^{\C}_{\tL,e})
\nonumber
\\
&+C^{(6)}_{ \textrm{SR}}
(\overline u_\tL  d_\tR)
(\overline e_{\tL} \nu^{\C}_{\tL,e})
+ C^{(6)}_{ \textrm{SL}}
(\overline u_\tR  d_\tL)
(\overline e_{\tL} \nu^{\C}_{\tL,e})
+ C^{(6)}_{ \textrm{T}} 
(\overline u_\tL \sigma^{\mu\nu} d_\tR)
(\overline e_{\tL}\sigma_{\mu\nu}\nu^{\C}_{\tL,e}) 
\Big] +{\rm h.c.}\;,  
\\  
\mathcal L^{(7)}_{\Delta L = 2} & = 
\frac{\sqrt{2} G_F}{v} 
\Big[ C^{(7)}_{\textrm{VL}}
(\overline u_\tL \gamma^\mu d_\tL)
(\overline e_{\tL}i\overleftrightarrow{\partial_\mu}
\nu^{\C}_{\tL,e})
+
C^{(7)}_{\textrm{VR}}
(\overline u_\tR \gamma^\mu d_\tR)
(\overline e_{\tL}i \overleftrightarrow{\partial_\mu}
\nu^{\C}_{\tL,e}) 
\Big]  + {\rm h.c.}\;,
\\
\mathcal L^{(9)}_{\Delta L = 2} & = 
\frac{1}{v^5}
\Big[C_{1\textrm{L}}^{(9)}
(\overline{u}_\tL  \gamma_\mu  d_\tL) (\overline{u}_\tL  \gamma^\mu  d_\tL)
(\overline{e_\tL}e_\tL^\C)
+C_{4\textrm{L}}^{(9)}
(\overline{u}_\tL  \gamma_\mu  d_\tL) (\overline{u}_\tR  \gamma^\mu  d_\tR)
(\overline{e_\tL}e_\tL^\C)
\Big]
+{\rm h.c.}\;,
\end{align}
\end{subequations}
where $u_{\tL,\tR}$, $d_{\tL,\tR}$, and $e_{\tL,\tR}$ denote the left- and right-handed up quark, down quark, and electron fields in the mass eigenstate, respectively, and $\nu_{\tL,e}$ represents the electron-type neutrino in the flavor eigenstate. 
We have neglected the dim-8 Lagrangian terms since their contributions are suppressed relative to those from dim-7 long-distance interactions by a factor of $p/v\sim\calO(10^{-4})$.
The coefficients in \cref{eq:LEFT0vbb} are matched at $\Lambda_{\tt EW}$ with the SMEFT interactions listed in \cref{tab:dim7_SMEFT_OP} by integrating out the heavy SM particles ($W,Z,h,t$) after electroweak symmetry breaking. 
The matching results are summarized in \cref{tab:matchingSMEFT2LEFT} for both the up- and down-quark flavor bases.
These results differ from those obtained earlier in 
Refs.~\cite{Cirigliano:2017djv,Liao:2020zyx} due to a more detailed treatment of the quark flavor basis here. While Ref.\,\cite{Cirigliano:2017djv} performs the matching without including CKM matrix elements before integrating out the $W^\pm$ bosons, Ref.\,\cite{Liao:2020zyx} includes CKM effects only in the up-quark flavor basis.

%%%%%
\begin{table}[h]
\center
\resizebox{\linewidth}{!}{
\renewcommand{\arraystretch}{1}
\begin{tblr}{
	colspec={|c | c l l c |},
	vline{1,2,4,6} = {1}{solid},
	vline{2,5} = {2-11}{2pt,solid,cyan},
	vline{3,6} = {5-14}{2pt,solid,purple},
	vline{1} = {1-14}{solid},
	vline{4} = {2-14}{dashed},
	}
\hline
\bf{\large Basis}
&
&~~\bf{ LEFT operators} 
&~~\bf{ Matching results at electroweak scale $\Lambda_{\tt EW}$} &
\\
\cline{1-1}\cline[1.5pt,cyan]{2-4}\cline{5-5}
{\color{cyan} Up} & 
& $\calO_{\textrm{SL}}^{(6)} = \sqrt{2}G_F(\overline{u_\tR} d_\tL)(\overline{e_\tL}\nu^{\C}_{\tL,e})$
& $C_{\textrm{SL}}^{(6)} = v^3\Big({\frac{1}{\sqrt{2}}} V_{w1} C_{\overline{Q}uLLH}^{w111*} + {\frac{1}{2v}} m_u V_{ud} C_{DLDH2}^{11*}\Big)$ &
\\ 
{\color{cyan} flavor} & 
& $ \calO_{\textrm{SR}}^{(6)} = \sqrt{2}G_F(\overline{u_\tL} d_\tR)(\overline{e_\tL}\nu^{\C}_{\tL,e})$ 
& $C_{\textrm{SR}}^{(6)} = v^3\Big(\frac{1}{2\sqrt{2}}C_{\overline{d}LQLH1}^{1111*} - {\frac{1}{2v}} m_d V_{ud} C_{DLDH2}^{11*}\Big)$ &
\\ 
{\color{cyan}  basis} & 
& $\calO_{\textrm{T}}^{(6)} = \sqrt{2}G_F(\overline{u_\tL} \sigma^{\mu\nu} d_\tR)(\overline{e_\tL}\sigma_{\mu\nu}\nu^{\C}_{\tL,e})$ 
& $C_{\textrm{T}}^{(6)} = v^3\Big(\frac{1}{8\sqrt{2}} C_{\overline{d}LQLH1}^{1111*} +\frac{1}{4\sqrt{2}}C_{\overline{d}LQLH2}^{1111*} \Big)$ &
\\
\cline[1.5pt,purple]{3-5}%%%
& & $\calO_{m_{\beta\beta}} = -\frac{1}{2}m_{\beta\beta}\overline{\nu^{\C}_{\tL,e}}\nu_{L,e}$ 
& $m_{\beta\beta} = -v^2C_{LH5}^{11}-\frac{1}{2} v^4 C_{LH}^{11}$ &
\\ 
& & $\calO_{\textrm{VL}}^{(6)} = \sqrt{2}G_F(\overline{u_\tL} \gamma^\mu d_\tL)(\overline{e_\tR}\gamma_\mu\nu^{\C}_{\tL,e})$ 
& $C_{\textrm{VL}}^{(6)} = v^3V_{ud}\Big(-\frac{1}{\sqrt{2}}C_{LeDH}^{11*} 
+ 4\frac{m_e}{v}g^{-1}C_{LHW}^{11*} \Big)$ &
\\
& & $\calO_{\textrm{VR}}^{(6)} = \sqrt{2}G_F(\overline{u_\tR} \gamma^\mu d_\tR)(\overline{e_\tR}\gamma_\mu\nu^{\C}_{\tL,e})$ 
& $C_{\textrm{VR}}^{(6)}= \frac{v^3}{2\sqrt{2}}C_{\overline{d}LueH}^{1111*}$ &
\\
& &$\calO_{\textrm{VL}}^{(7)} = \frac{\sqrt{2}}{v}G_F(\overline{u_\tL}\gamma^\mu d_\tL)(\overline{e_\tL} \overleftrightarrow{\partial_\mu}\nu^{\C}_{\tL,e})$ 
& $C_{\textrm{VL}}^{(7)} = -v^3V_{ud}\Big(C_{DLDH1}^{11*}+ {\frac{1}{2}}  C_{DLDH2}^{11*} +4g^{-1} C_{LHW}^{11*}\Big)$ &
\\
& & $\calO_{\textrm{VR}}^{(7)} = \frac{\sqrt{2}}{v}G_F(\overline{u_\tR}\gamma^\mu d_\tR)(\overline{e_\tL} \overleftrightarrow{\partial_\mu}\nu^{\C}_{\tL,e})$ 
& $C_{\textrm{VR}}^{(7)} = -v^3C_{\overline{d}uLDL}^{1111*}$ &
\\
& & $\calO_{\textrm{1L}}^{(9)} = \frac{1}{v^5}(\overline{u_\tL}\gamma_\mu d_\tL) (\overline{u_\tL}\gamma^\mu d_\tL) (\overline{e_\tL} e_\tL^\C) $ 
& $C_{\textrm{1L}}^{(9)} = -v^3V_{ud}^2\big(2C_{DLDH1}^{11*}+8g^{-1}C_{LHW}^{11*} \big)$ &
\\
& & $\calO_{\textrm{4L}}^{(9)} = \frac{1}{v^5}(\overline{u_\tL}\gamma_\mu d_\tL) (\overline{u_\tR}\gamma^\mu d_\tR) (\overline{e_\tL} e_\tL^\C) $ 
& $C_{\textrm{4L}}^{(9)} = - 2v^3 V_{ud} C_{\overline{d}uLDL}^{1111*}$ &
\\
\cline[1.5pt,cyan]{2-4}%%%
{\color{purple} Down} & 
& $\calO_{\textrm{SL}}^{(6)} = \sqrt{2}G_F(\overline{u_\tR} d_\tL)(\overline{e_\tL}\nu^{\C}_{\tL,e})$ 
& $C_{\textrm{SL}}^{(6)} = v^3\Big({\frac{1}{\sqrt{2}} } C_{\overline{Q}uLLH}^{1111*} + {\frac{1}{2v}} m_u V_{ud} C_{DLDH2}^{11*}\Big)$ &
\\ 
{\color{purple} flavor} & 
&$\calO_{\textrm{SR}}^{(6)} = \sqrt{2}G_F(\overline{u_\tL} d_\tR)(\overline{e_\tL}\nu^{\C}_{\tL,e})$ 
& $C_{\textrm{SR}}^{(6)} = v^3\Big(\frac{1}{2\sqrt{2}} V_{1w} C_{\overline{d}LQLH1}^{11w1*}- {\frac{1}{2v}} m_d V_{ud} C_{DLDH2}^{11*}\Big)$ &
\\
{\color{purple} basis }& 
& $\calO_{\textrm{T}}^{(6)} = \sqrt{2}G_F(\overline{u_\tL} \sigma^{\mu\nu} d_\tR)(\overline{e_\tL}\sigma_{\mu\nu}\nu^{\C}_{\tL,e})$ 
& $C_{\textrm{T}}^{(6)} = v^3 V_{1w} \Big(\frac{1}{8\sqrt{2}} C_{\overline{d}LQLH1}^{11w1*}+\frac{1}{4\sqrt{2}}C_{\overline{d}LQLH2}^{11w1*}\Big)$ &
\\
\cline{1-2}\cline[1.5pt,purple]{3-6}
\end{tblr} }
\caption{Matching relations between the SMEFT interactions in \cref{tab:dim7_SMEFT_OP} and the LEFT operators adopted in Refs.~\cite{Cirigliano:2017djv,Cirigliano:2018yza,Scholer:2023bnn}.}
\label{tab:matchingSMEFT2LEFT}
\end{table}
%%%%%

After accounting for the QCD running effect of the above LEFT interactions in \cref{eq:LEFT0vbb} 
from $\Lambda_{\tt EW}$ down to the chiral symmetry breaking scale ($\Lambda_\chi \sim \text{GeV}$), 
these interactions are further matched onto hadronic LNV interactions within the chiral EFT ($\chi$EFT) framework, where the relevant hadronic degrees of freedom are the nucleons and pions. 
Based on those $\chi$EFT interactions, the $0\nu\beta\beta$ decay transition operators can be constructed, which are then used to formulate the relevant nuclear matrix elements and decay amplitudes.
The half-life of $0\nu\beta\beta$ is obtained from the decay amplitude squared by incorporating the phase space factors. 
All of these steps have been implemented in the \texttt{$\nu$DoBe} package \cite{Scholer:2023bnn}, whose default nuclear input parameters are to be adopted in our numerical analysis. 
			
%%%%%%%%%%%%%%%%%%%%%%%%%%%%%%%%%%%%%%%%
\subsection{Interfacing \texttt{D7RGESolver} with \tf{\texttt{$\nu$DoBe}}{vDoBe}}
%%%%%%%%%%%%%%%%%%%%%%%%%%%%%%%%%%%%%%%%

In this work, we calculate the half-life of $0\nu\beta\beta$ decay using the \texttt{$\nu$DoBe} package~\cite{Scholer:2023bnn}, which automatically computes the half-life once the WCs of the relevant SMEFT or LEFT operators at
$\Lambda_{\texttt{EW}}$ are provided. We first compute the RGE evolution of the dim-5 and dim-7 SMEFT operators from the NP scale down to the electroweak scale using \texttt{D7RGESolver}, as illustrated in \cref{sec:D7rgesolver}. Subsequently, we match the WCs of SMEFT to the corresponding LEFT operators at $\Lambda_{\texttt{EW}}$ according to \cref{tab:matchingSMEFT2LEFT},
and input these WCs of LEFT into \texttt{$\nu$DoBe} to calculate the half-life of $0\nu\beta\beta$ decay. These steps can be accomplished by the following commands:
%%%%%%%%%%%%
\begin{lstlisting}[language=iPython, numbers=none, 
breaklines=true, 
xleftmargin=2.5em, 
xrightmargin=0.5em]
C_in = {"dLQLH1_1121": 2.77e-16+0j}     #Input the WCs at NP scale 
scale_in = 1e4                          #Input energy scale in units of GeV
scale_out = 80                          #Output energy scale in units of GeV
C_out = solve_rge(scale_in,scale_out,C_in,basis="down",method="integrate")
LEFT_WCs=extract_0vbb_LEFT(C_out,basis="down") #Match SMEFT to LEFT at EW scale
model = LEFT(LEFT_WCs)  #Input the LEFT at EW scale into nudobe
model.t_half("136Xe")   #Calculate the half-life via nudobe, in units of year
\end{lstlisting}
%%%%%%%%%%%%%
with the output half-life being
{\small\begin{verbatim}
2.2967762685227545e+26
\end{verbatim}}
\normalsize
in units of year.
The function \texttt{extract\_0vbb\_LEFT}, defined in \texttt{ovbb.py}, is used to match the WCs of the SMEFT operators to the LEFT operators according to \cref{tab:matchingSMEFT2LEFT}. The last two lines are \texttt{$\nu$DoBe} commands that are used to calculate the half-life of \( 0\nu\beta\beta \) decay for a given nucleus. Based on the RGEs solved by \texttt{D7RGESolver} and the half-life results obtained from \texttt{$\nu$DoBe}, we define several functions in \texttt{ovbb.py} to derive limits on the dim-5 and dim-7 SMEFT operators at any scale above \( \Lambda_{\tt{EW}} \) from \( 0\nu\beta\beta \) decay. Here, we take the function \texttt{find\_0vbb\_limit} as an example. This function is used for constraining one single operator a time. The user can input the following commands: 
\begin{lstlisting}[language=iPython, numbers=none, 
breaklines=true, 
xleftmargin=2.5em, 
xrightmargin=0.5em]
from ovbb import find_0vbb_limit
SMEFT_operators_0vbb = ["LH5_11", "LH_11"]
find_0vbb_limit( 
    operators=SMEFT_operators_0vbb,     # The operators to scan
    scale=1e4,                          # The NP scale of the operators
    basis="down",                       # The quark-flavor basis
    target_half_life=2.3e26,            # The upper limit of the half-life
    isotope="136Xe")                    # The isotope used in 0vbb experiments
\end{lstlisting}
\noindent with the output being:
{\footnotesize
\begin{verbatim}
Limit on WC LH5_11 at 1.000e+04 GeV is 5.606e-16 GeV^4-d, Corresponding half-life = 2.300e+26 yr
Limit on WC LH_11 at 1.000e+04 GeV is 1.923e-20 GeV^4-d, Corresponding half-life = 2.300e+26 yr

=== Summary Table ===
+--------+------------------+-------------------------------------+
|        |   Scale_in (GeV) |   Limit on WCs ($\text{GeV}^{4-d}$) |
+========+==================+=====================================+
| LH5_11 |            10000 |                           5.606e-16 |
+--------+------------------+-------------------------------------+
| LH_11  |            10000 |                           1.923e-20 |
+--------+------------------+-------------------------------------+
\end{verbatim}}
\normalsize
\noindent 
By using the function \texttt{find\_0vbb\_limit}, we obtain the constraints on the WCs of these SMEFT operators at $\Lambda = 10\,\rm TeV$, assuming single-operator dominance. These constraints are listed in the third column of the ``Summary Table''. For detailed usage, please refer to the example provided in the \texttt{example} folder.

%%%%%%%%%%%%%%%%%%%%%%%%%%%%%%%%%%%%%%%%%%%%%%%%%%%%%%%%%%%%%%%%
\subsection{RGE-improved analyses on dim-7 SMEFT interactions from \tf{$0\nu\beta\beta$}{0vbb} decay}
\label{sec:numerical}
%%%%%%%%%%%%%%%%%%%%%%%%%%%%%%%%%%%%%%%%%%%%%%%%%%%%%%%%%%%%%%%%

Since the recent KamLAND-Zen experiment has provided the hitherto most stringent limit on the $0\nu\beta\beta$ decay half-life of the ${}^{136}$Xe nucleus, with $T_{1/2}^{0\nu}(^{136}\text{Xe})>2.3\times10^{26}$ yr \cite{KamLAND-Zen:2022tow}, we will exclusively use this bound to constrain the dim-7 SMEFT interactions, assuming one operator is active at a time. Note that we treat all SMEFT dim-5 and dim-7 WCs as free parameters and constrain them using the $0\nu\beta\beta$ decay, without attempting to reproduce the neutrino oscillation data. Since the neutrino mass matrix similarly receives multiple contributions in the SMEFT framework, this can always be achieved with ample parameter space.

%%%%%%%%%%%%%%%%%%%%%%%%%%%%%%%%%%%%%%%%%%%%%%%%%%%%%%%%%%%%%%%%
\subsubsection{RGE-mixed contributions to neutrino mass operators}
\label{sec:numassmixing}
%%%%%%%%%%%%%%%%%%%%%%%%%%%%%%%%%%%%%%%%%%%%%%%%%%%%%%%%%%%%%%%%

\begin{figure}[t] 
\centering
\includegraphics[width=0.8\linewidth]{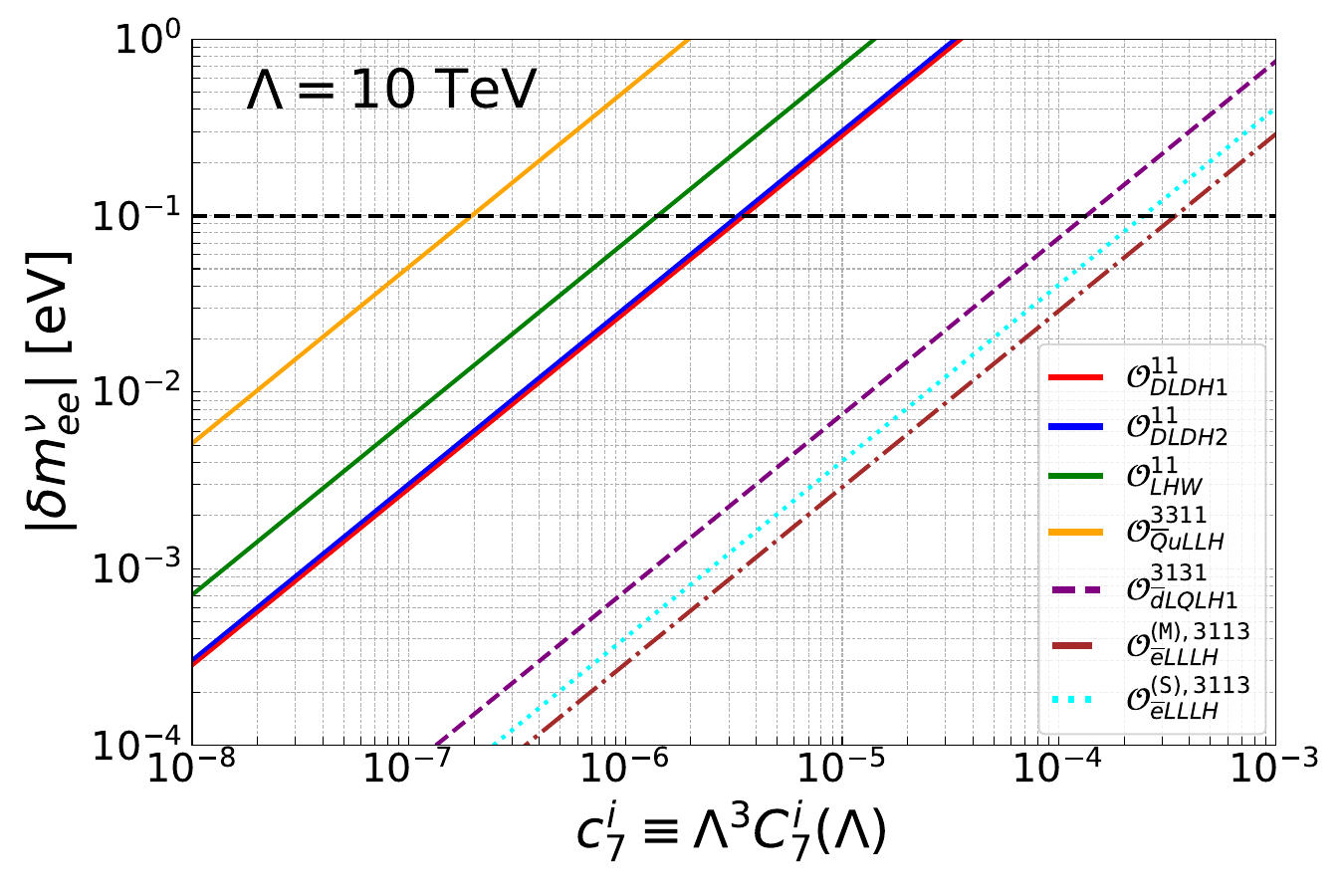}
\caption{
The correction to the effective electron-type neutrino mass $m^{\nu}_{ee}$ due to the RGE running contribution from certain dim-7 SMEFT operators.  The horizontal dashed line denotes a typical upper limit $m_{ee}^{\nu}\lesssim 0.1~\rm eV$.}
\label{fig:numass_dim7_mixing}
\end{figure}

There are three mechanisms contributing to $0\nu\beta\beta$ decay: the mass, the long-distance, and the short-distance mechanisms. The mass mechanism arises from the insertion of the Majorana neutrino mass term. The long-distance mechanism is mediated by the exchange of a light neutrino between a standard dim-6 $\beta$-decay interaction and a dim-6/7 LNV interaction in the LEFT, while the short-distance contribution is induced by contact dim-9 LEFT interactions. 
Among all SMEFT operators up to dim 7, the neutrino mass operators $\calO_{LH5}^{11}$ and $\calO_{LH}^{11}$ receive the most stringent constraints from $0\nu\beta\beta$ decay with the corresponding LNV scale $\Lambda>\calO(10^{12}\,\rm TeV)$  and $\Lambda>\calO(10^{4}\,\rm TeV)$, respectively. 
This makes other dim-7 operators that mix with $\calO_{LH5}^{11}$ and $\calO_{LH}^{11}$ under RGE evolution also receive more stringent constraints than those from their direct tree-level contributions to the process. For instance, 
let us consider the RGE running contribution to the neutrino mass term from the following dim-7 operators: 
$\calO_{DLDH1}^{11}$, $\calO_{DLDH2}^{11}$, $\calO_{LHW}^{11}$, $\calO_{\overline{Q}uLLH}^{3311}$, $\calO_{\overline{e}LLLH}^{{\tt (S)},3113}$, $\calO_{\overline{e}LLLH}^{{\tt (M)},3113}$, and $ \calO_{\overline{d}LQLH1}^{3131}$. 
\cref{fig:numass_dim7_mixing} shows the corrections to the effective electron-type neutrino mass $m_{ee}^\nu$
as a function of their dimensionless WC $c_i$, after incorporating the RGE effect solved by \texttt{D7RGESolver} with a fixed NP scale $\Lambda=10\,\rm TeV$.  
As observed from the plot, these operators can induce a correction to $m_{ee}^\nu$ that is comparable to the current experimental limit on $m_{ee}^\nu$ for $c_i\sim 10^{-7}-10^{-4}$, depending on the specific operator. Notably, the operator $\calO_{\overline{Q}uLLH}^{3311}$, which involves both left- and right-handed third-generation up-type quarks, provides the largest correction to $m_{ee}^\nu$ for similar values of $c_i$. This is due to the large top Yukawa coupling $y_t\sim\calO(1)$ involved in the RGE mixing.
Consequently, this operator is subject to a particularly strong constraint from $0\nu\beta\beta$ decay.    
Note that the operators $\calO_{\overline{Q}uLLH}^{3311}$ and 
$\calO_{\overline{d}LQLH1}^{{\tt (M)},3131}$ each involve only quark fields of the same flavor. As a result, their contributions to neutrino masses exhibit only a very weak dependence on the choice of the quark flavor basis in the RGE running calculation, which can be seen from the RGEs of $\calO_{LH5}$ and $\calO_{LH}$ presented in \cref{eq:RGE_Weinberg_dim5,eq:RGE_Weinberg_dim7}.
Therefore, we have neglected this minor difference in the above plot.
		
\subsubsection{RGE-improved constraints on operators involving only first-generation fermions}
\label{sec:1_gene_ope}

In this part, we examine the RGE effects on the 11 dim-5 and dim-7 SMEFT operators that involve only first-generation fermions. Their contributions to $0\nu\beta\beta$ decay have been studied in Refs.~\cite{Cirigliano:2017djv,Liao:2019tep, Scholer:2023bnn}, in which only a subset of RGEs are considered. We will revisit these constraints by incorporating the full SMEFT RGE effects. 
As these operators involve only first-generation quarks, the constraints are insensitive to the choice of quark flavor basis, and we therefore neglect the minor differences below.

For the dim-5 Weinberg operator $\mathcal{O}_{LH5}^{11}$, including RGE running slightly weakens the constraint on its WC at a NP scale. Specifically, we find the limit 
$|C_{LH5}^{11}(\Lambda)| > 5.61 \times 10^{-16}\,\text{GeV}^{-1}$ at a NP scale of $\Lambda = 10$ TeV, compared to the constraint at the electroweak scale, $|C_{LH5}^{11}(\Lambda_{\texttt{EW}})| > 4.99 \times 10^{-16}\,\text{GeV}^{-1}$, derived from $0\nu\beta\beta$ decay.

\begin{figure}[t] 
\centering
\includegraphics[width=0.98\linewidth]{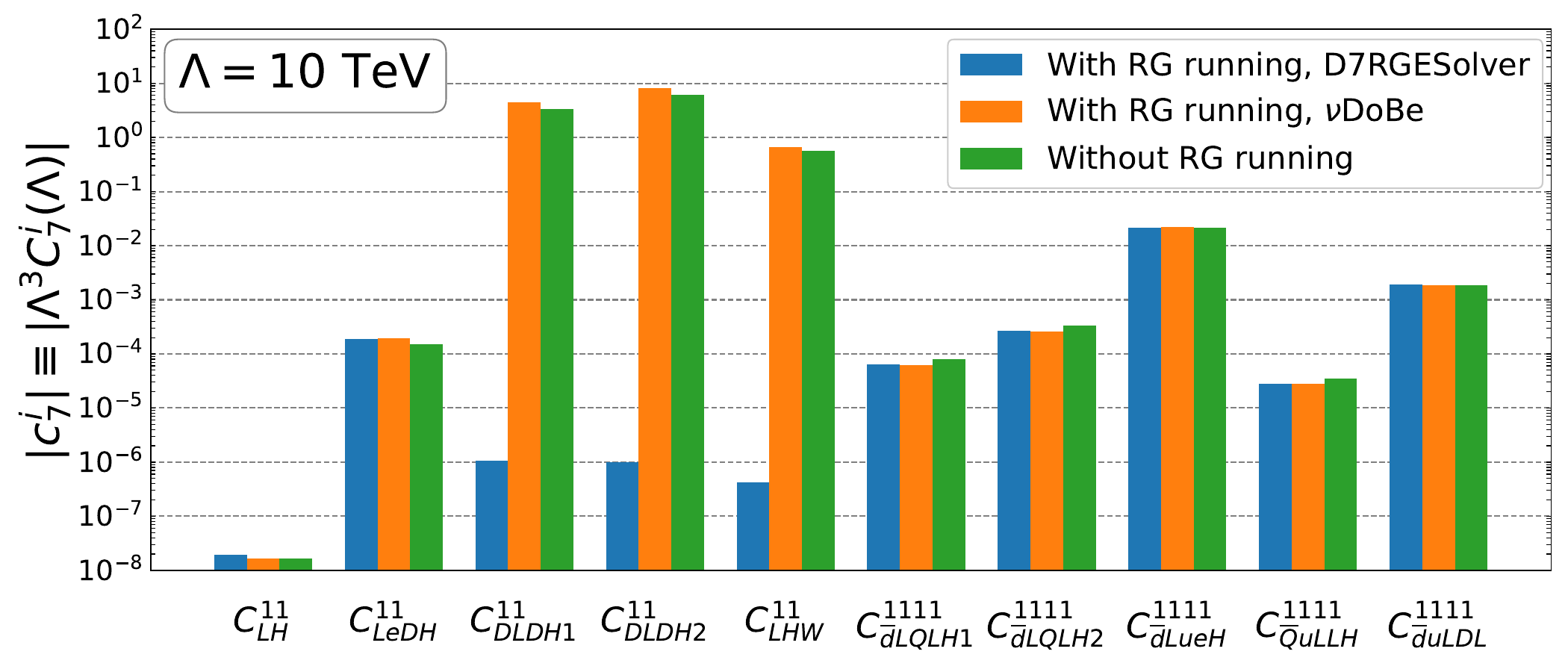}
\caption{
Comparison of constraints on dimensionless WCs $c_7^i$ for dim-7 LNV interactions involving only first-generation fermions at the NP scale $\Lambda = 10\,\rm TeV$,  with full, partial, or no RGE effects as indicated by the blue, orange, and green colors, respectively.}
\label{fig:0vbb_limit_single_Ops}
\end{figure}

In \cref{fig:0vbb_limit_single_Ops}, we present the constraints on the dimensionless WCs $c_7^i$ of these dim-7 first-generation-fermion operators at the NP scale $\Lambda=10\,\rm TeV$ in three scenarios. 
The blue and orange bars show the results after including the RGE evolution of these operators from $\Lambda=10\,\rm TeV$ down to the electroweak scale $\Lambda_{{\tt EW}}$, 
calculated by using \texttt{D7RGESolver} and \texttt{$\nu$DoBe}, respectively, 
while the green bars represent results without accounting for the RGE effects. 
As clearly shown in the plot, the full and precise RGE evolution of these operators has a significant impact on their numerical constraints. In particular, the RGE-improved constraints on the operators $\calO_{LHW}^{11}$, $\calO_{DLDH1}^{11}$, and $\calO_{DLDH2}^{11}$ are stronger than those in the other two scenarios by a factor of $\calO(10^6)$. This enhancement in sensitivity arises from their contributions to the neutrino mass operators through RGE evolution as discussed in \cref{sec:numassmixing}. However, this crucial effect was not included in the $\nu \texttt{DoBe}$ package, which consequently yields a weaker bound.

\begin{figure}[t] 
\centering
\includegraphics[width=0.8\linewidth]{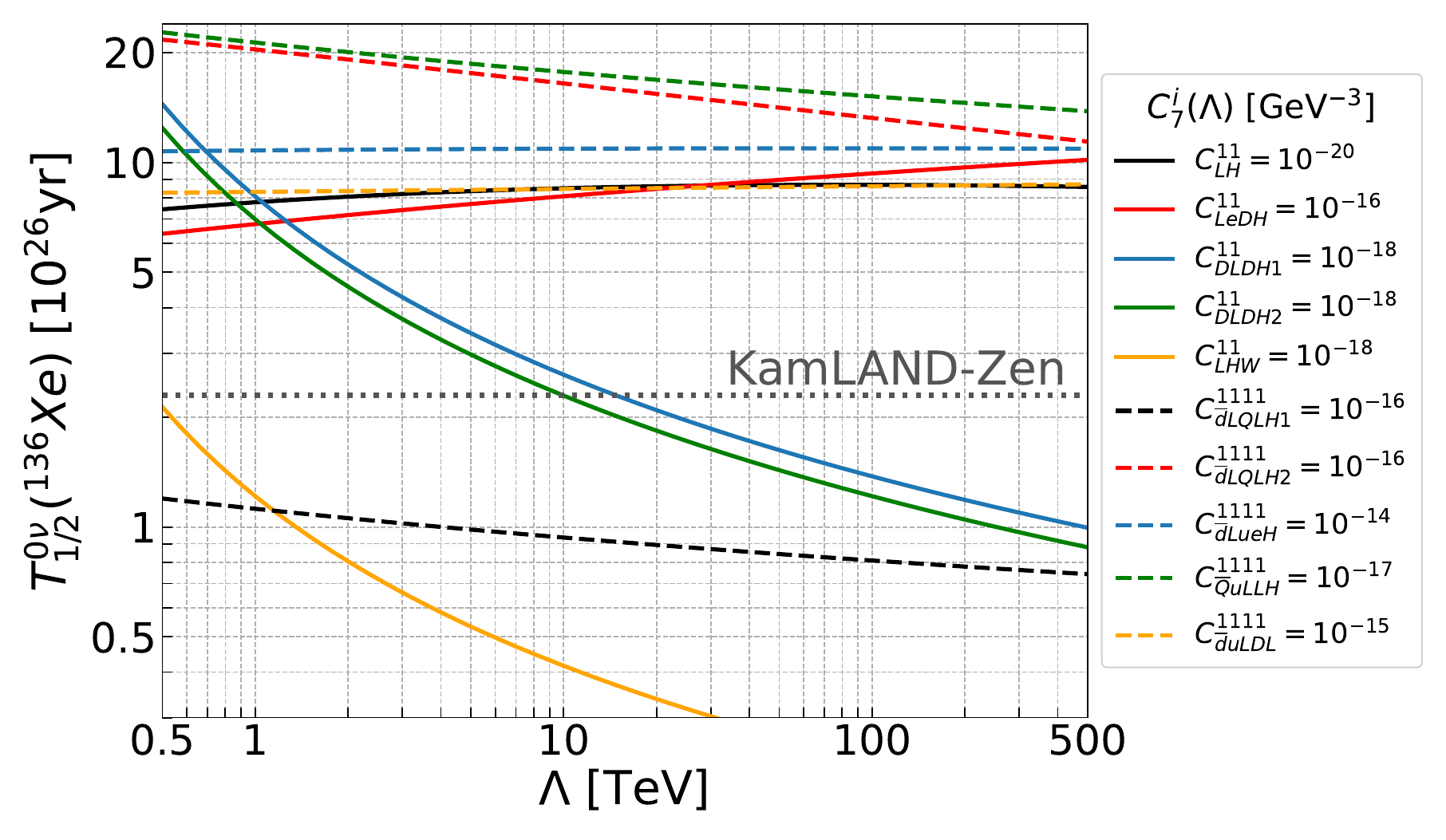}
\caption{
The half-life of $0\nu\beta\beta$ decay as a function of the NP scale $\Lambda$, assuming a fixed WC $C_7^i(\Lambda)$ for each dim-7 SMEFT operator at the NP scale.}
\label{fig:halflife_scan_scale}
\end{figure}

In \cref{fig:halflife_scan_scale}, we fix the WC of each dim-7 operator at an arbitrary NP scale $\Lambda$ and show the resulting half-life of $0\nu\beta\beta$ decay for the $\rm{}^{136}Xe$ nucleus as a function of $\Lambda$. The figure illustrates that the RGE effect plays a significant role in predicting the $0\nu\beta\beta$ decay half-life. This effect is especially pronounced for the operators $\mathcal{O}_{DLDH1,2}^{11}$ and $\mathcal{O}_{LHW}^{11}$, as their contributions to $0\nu\beta\beta$ decay arise primarily through RGE mixing into the neutrino mass operator. Consequently, the induced decay rates at $\Lambda = 500$ TeV can be an order of magnitude larger than those at $\Lambda = 0.5$ TeV. Moreover, RGE effects also play a substantial role in the contributions to $0\nu\beta\beta$ decay from $\mathcal{O}_{\bar{d}LQLH1,2}^{1111}$, $\mathcal{O}_{\bar{Q}uLLH}^{1111}$, and $\mathcal{O}_{LeDH}^{11}$, whereas the dependence of RGE evolution on the NP scale is less pronounced for the operators $\mathcal{O}_{LH}^{11}$, $\mathcal{O}_{\bar{d}LueH}^{1111}$, and $\mathcal{O}_{\bar{d}uLDL}^{1111}$.

\begin{figure}[t] 
\centering
\includegraphics[width=0.31\linewidth]{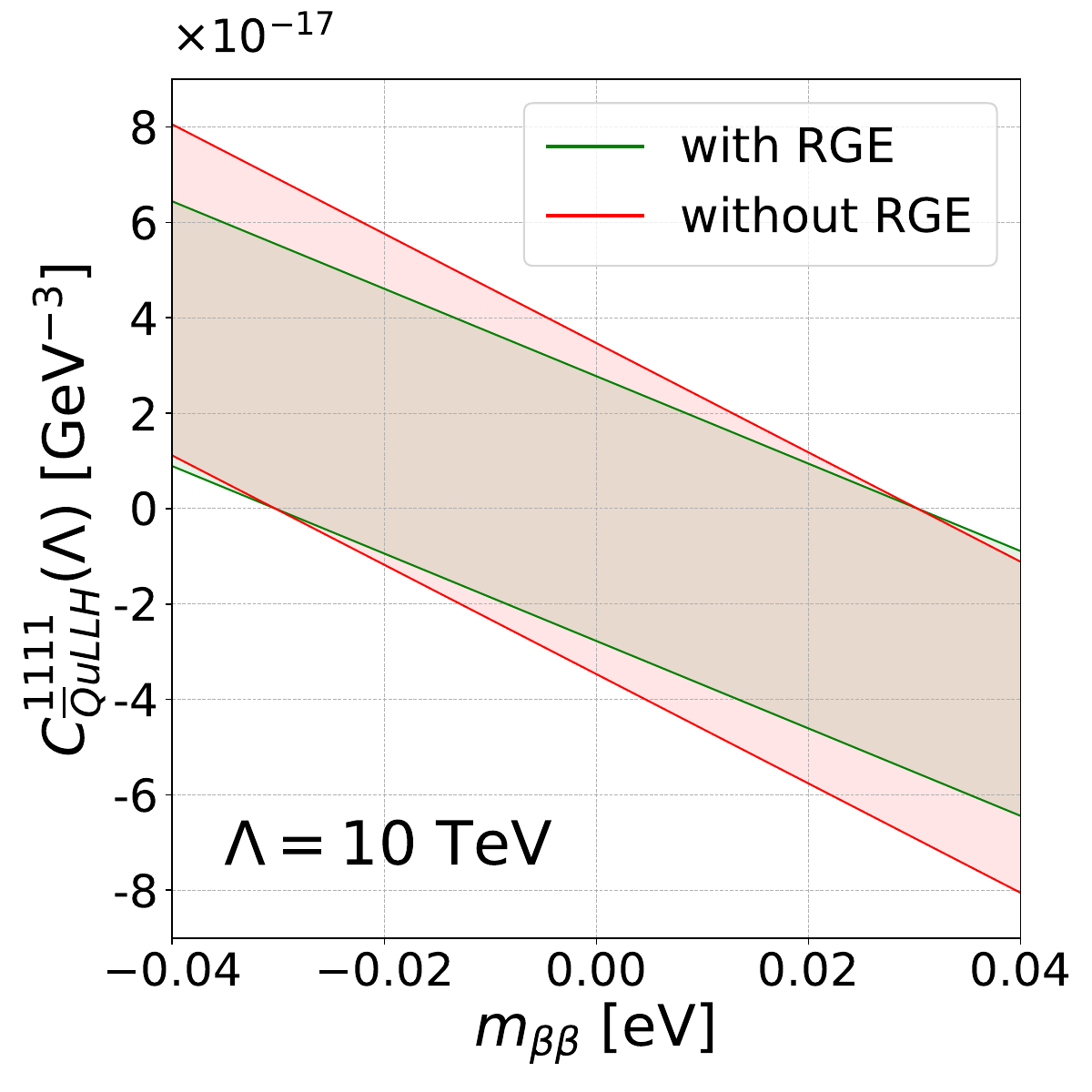}
\includegraphics[width=0.31\linewidth]{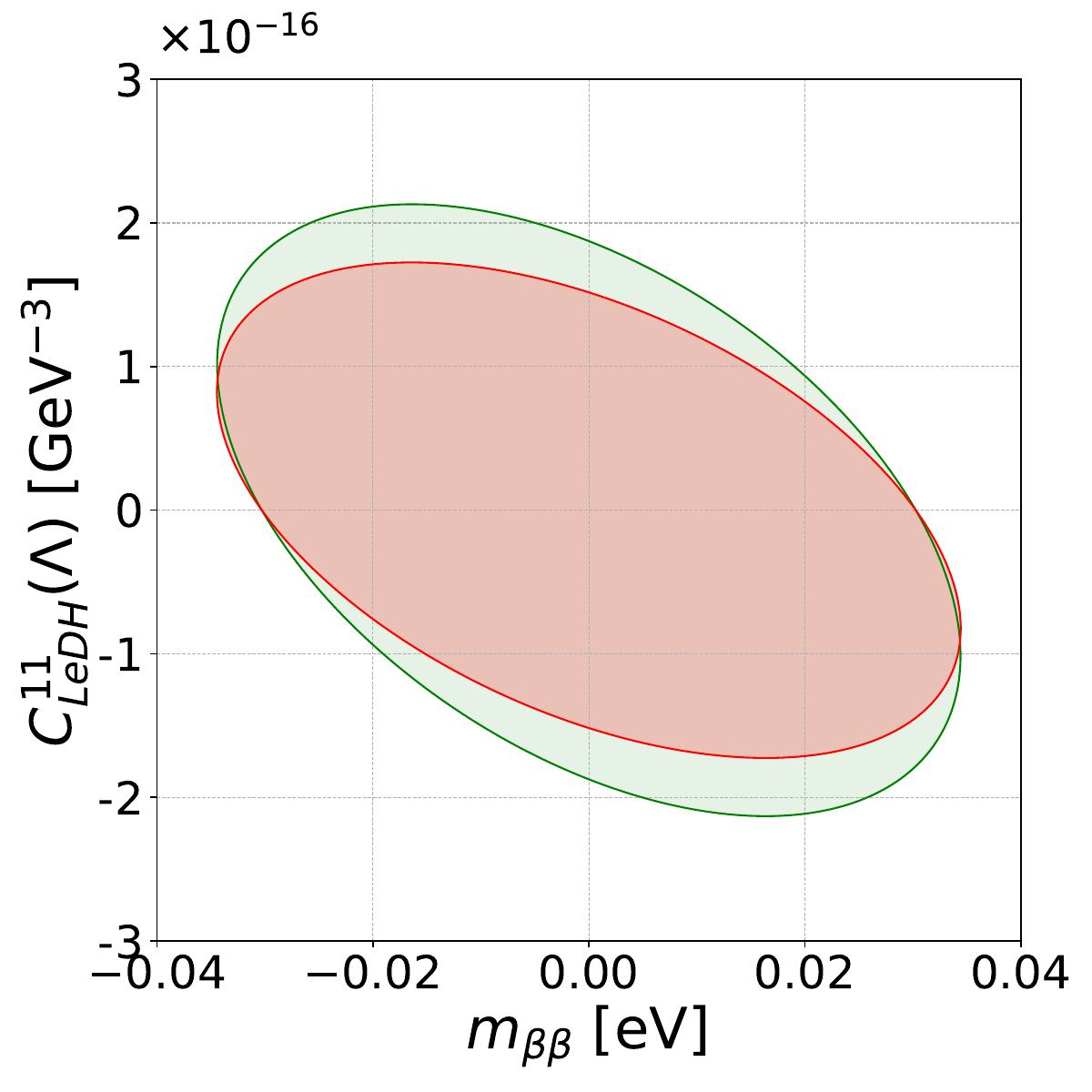} 
\includegraphics[width=0.31\linewidth]{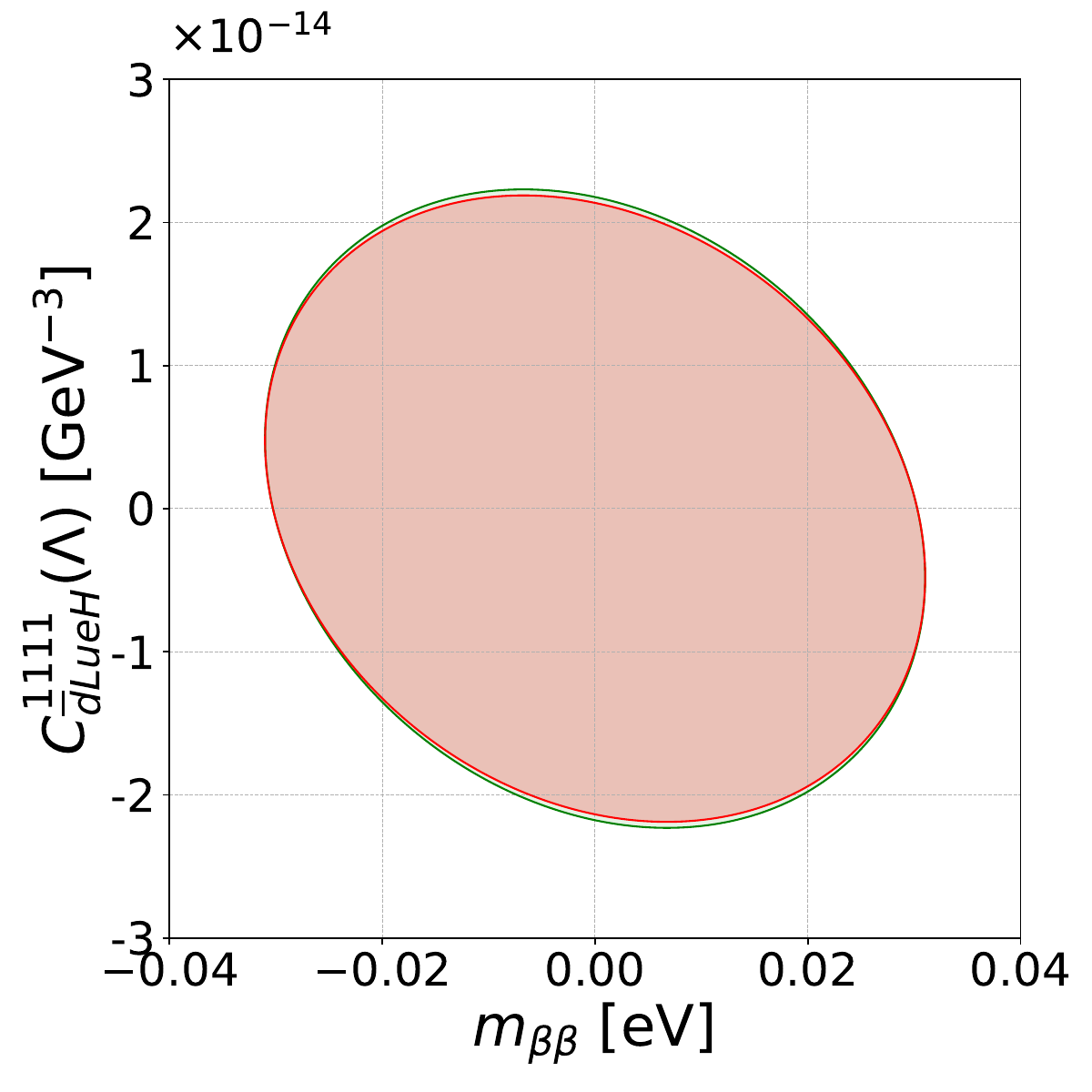}\\
\includegraphics[width=0.31\linewidth]{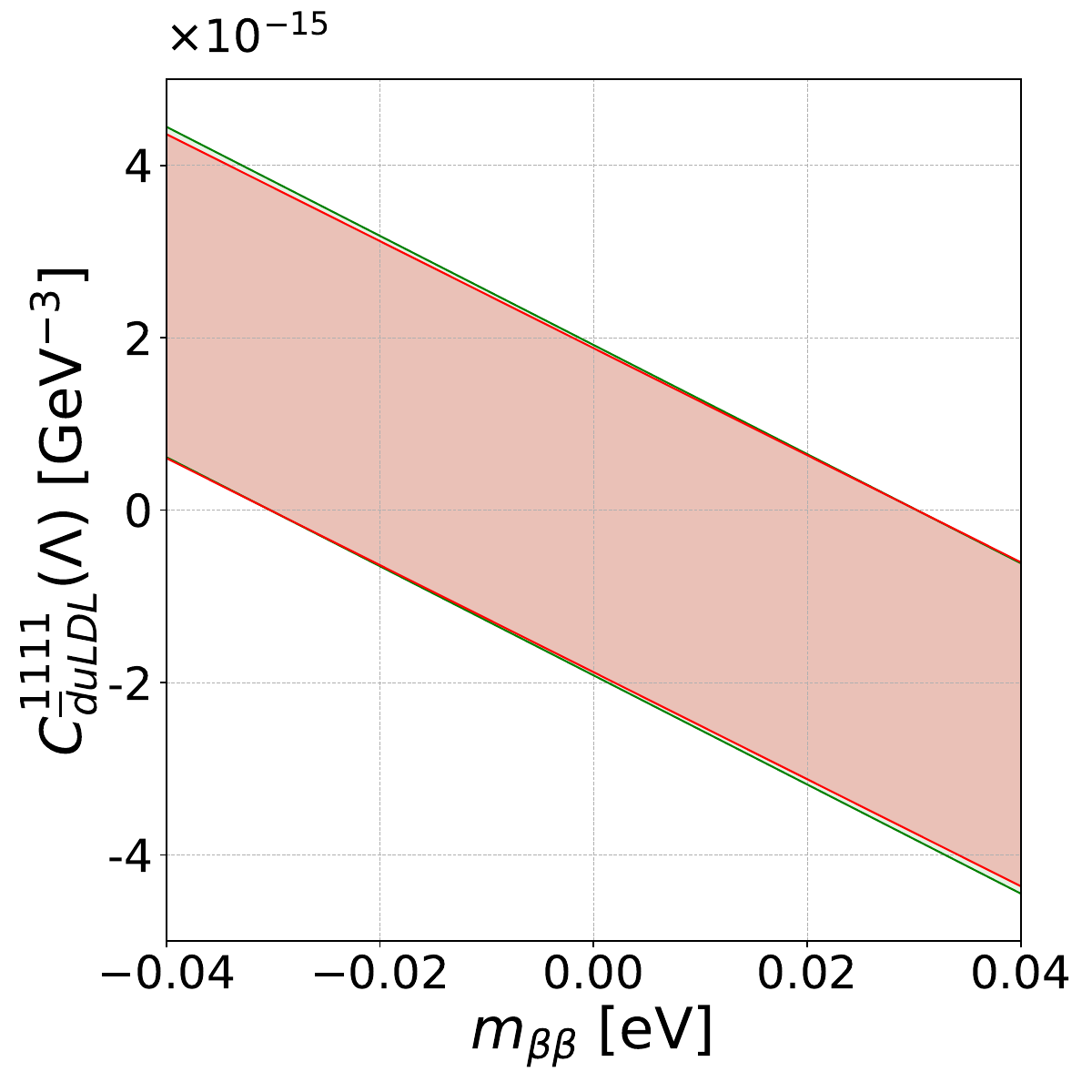}		
\includegraphics[width=0.31\linewidth]{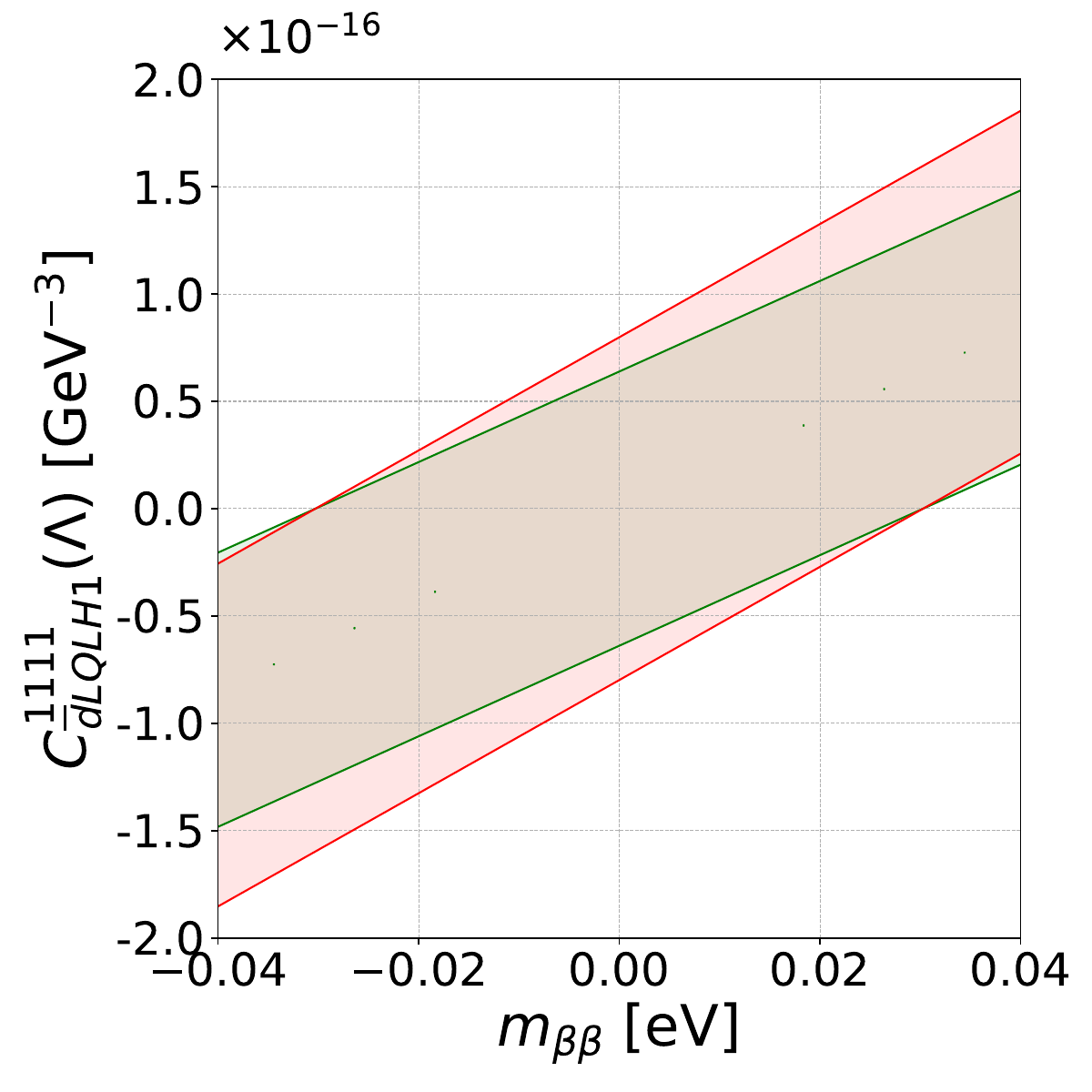}
\includegraphics[width=0.31\linewidth]{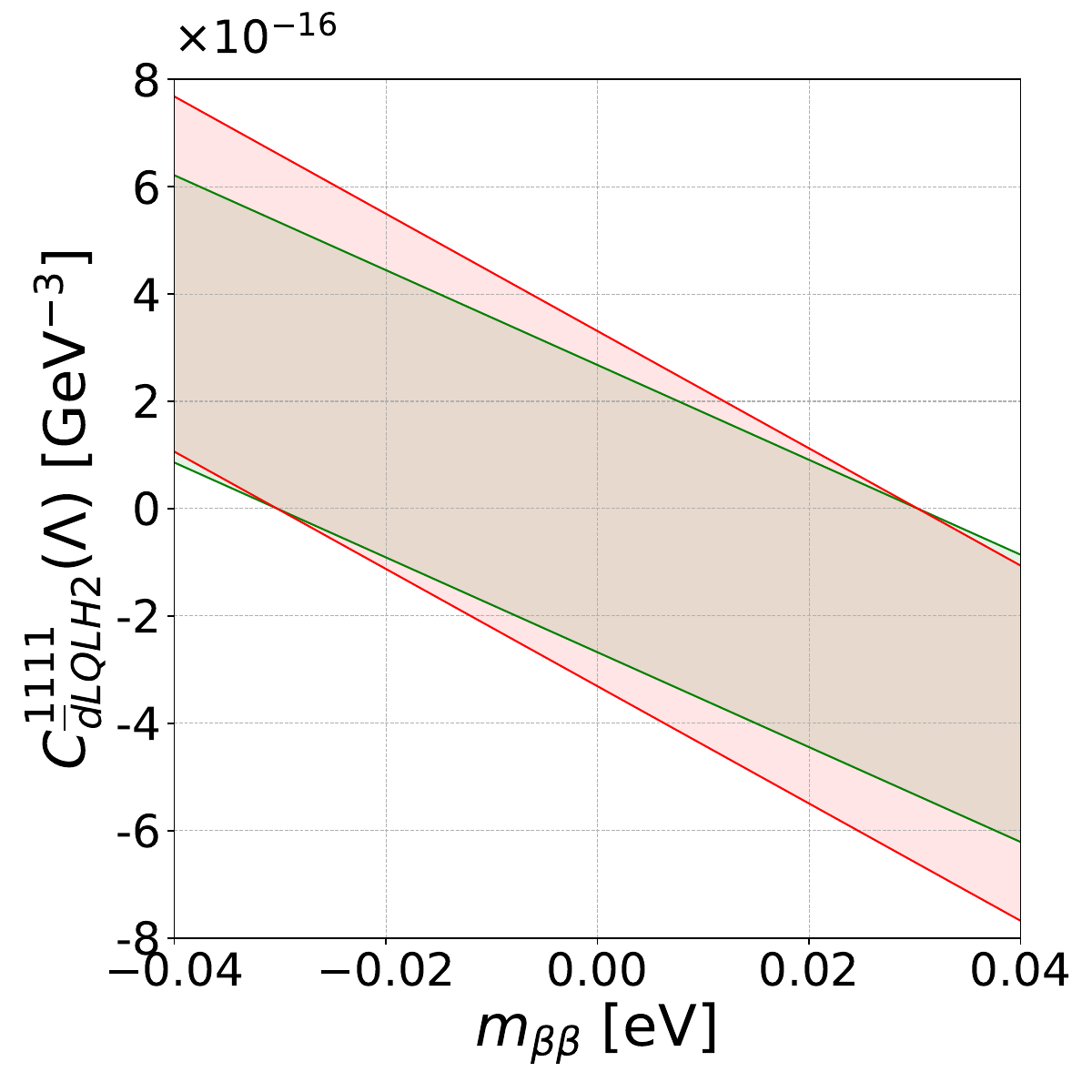}
\caption{
Constraints on the two-parameter space ($m_{\beta\beta}-C_7^i$ plane) derived from the current KamLAND-Zen experimental limits. The green (red) regions represent the allowed regions with (without) RGE effects.
}
\label{fig:2D_limit_C7_mbb}
\end{figure}	

It is a reasonable and practical assumption that the  $0\nu\beta\beta$ decay proceeds jointly via the standard mass mechanism ($m_{\beta\beta}$) and a dim-7 contribution ($C_7^i$) through either long-distance or short-distance mechanisms, as was studied in~\cite{Scholer:2023bnn,Chen:2024ctu,Ding:2024obt}. Following a similar approach, we consider the full RGE effects on the allowed parameter space in the two-dimensional plane defined by $m_{\beta\beta}-C_7^i$. 
For concreteness, we take the NP scale $\Lambda=10\,\rm TeV$ and consider the following 6 dim-7 operators involving only first-generation fermions: 
$\calO_{LeDH}^{11}$,
$\calO_{\overline{d}LueH}^{11}$, 
$\calO_{\overline{d}uLDL}^{1111}$, 
$\calO_{\overline{Q}uLLH}^{1111}$, 
$\calO_{dLQLH1}^{1111}$, and $\calO_{dLQLH2}^{1111}$. 
The mixing of these operators and the neutrino mass operators from RGE evolution is negligible.
The green and red regions in the $m_{\beta\beta}-C_7^i$ plane of  \cref{fig:2D_limit_C7_mbb} represent the allowed parameter space from the current KamLAND-Zen experiment, with and without RGE effects from $\Lambda = 10\,\rm TeV$ down to $\Lambda_{\tt EW}$, respectively. 
It can be seen that RGE running has a negligible effect on the parameter space of operators $\calO_{\overline{d}LueH}^{11}$ and $\calO_{\overline{d}uLDL}^{1111}$, but a noticeable impact on that of the remaining operators. 

The combined constraints on $m_{\beta\beta}$ and $C_7^i$ shown in 
\cref{fig:2D_limit_C7_mbb} appear as either a band or an ellipse, depending on whether the contribution from the dim-7 operator ${\cal O}_7^i$ interferes significantly with the standard mass mechanism from $m_{\beta\beta}$. Since the latter is associated with the left-handed electron field, the interference is significant if ${\cal O}_7^i$ also involves left-handed electron fields. This is the case for the operators ${\cal O}_{\overline{Q}uLLH}^{1111}$, ${\cal O}_{\overline{d}uLDL}^{1111}$, and ${\cal O}_{dLQLH1,2}^{1111}$. The constraint has a band structure since only the coherent sum of the two contributions can be bound. The correlation or anti-correlation between the two depends on their relative sign being negative or positive. On the opposite, if ${\cal O}_7^i$ involves exclusively right-handed electron fields, its interference with the standard mass mechanism is strongly suppressed by the electron mass. This will yield a decay rate that practically depends on a sum of $|C_7^i|^2$ and $m_{\beta\beta}^2$. This explains why the constraints on the WCs of the operators ${\cal O}_{LeDH}^{11}$ and ${\cal O}_{\overline{d}LueH}^{1111}$ are an ellipse.

Needless to say, our results depend on the choice of quantities with various uncertainties such as nuclear matrix elements, phase space factors, and strong low energy constants. But the qualitative features disclosed here are not affected by these uncertainties.

%%%%%%%%%%%%%%%%%%%%%%%%%%%%%%%%%%%%%%
\subsubsection{Complete RGE-improved constraints on dim-7 LNV interactions from \tf{$0\nu\beta\beta$}{0vbb}
}
%%%%%%%%%%%%%%%%%%%%%%%%%%%%%%%%%%%%%%

After discussing the general features arising from the RGE running and mixing in the previous parts, we now establish comprehensive constraints on all possible dim-7 operators with various lepton and quark flavor ombinations, taking into account the full set of RGEs. We fix the NP scale $\Lambda = 10$ TeV and assume that only one operator is active at this scale. We then run it down to the electroweak scale $\Lambda_{\tt EW}$ using the full RGEs solved by \texttt{D7RGESolver}, and calculate its contribution to the half-life of $0\nu\beta\beta$ decay, which can be expressed in a general form
$[T_{1/2}^{0\nu}]_{\tt SMEFT}(c_i)= \kappa_i |c_i|^2$,
where $c_i$ is the dimensionless WC of the inserted operator, and $\kappa$ is a numerical factor that incorporates both RGE running and nuclear effects.
From this, we derive a bound on $c_i$ by requiring $[T_{1/2}^{0\nu}]_{\tt SMEFT}(c_i) \gtrsim [T_{1/2}^{0\nu}]_{\tt KamLAND}^{90\,\%}=2.3\times10^{26}\,\rm yr$.   

Since the solution of the RGEs depends on the specific form of the SM Yukawa coupling matrices ($Y_{l}, Y_{u}$ and $Y_d$), 
we consider two popular choices of the quark flavor bases, the up-quark and down-quark flavor bases as detailed in \cref{sec:RGEdim7}, to present our numerical results. Additionally, the matching conditions between the SMEFT and LEFT interactions relevant to $0\nu\beta\beta$ also depend on the chosen quark flavor basis. We have accordingly incorporated the matching relations for both bases from \cref{tab:matchingSMEFT2LEFT} into \texttt{ovbb.py} to ensure consistent numerical evaluation. Our final results are presented in \cref{tab:limit_dim7_Ops}. The first and second columns present the results that are identical in both quark flavor bases, while the last three show results that differ between the two bases. 
From the table, we observe that those operators involving the second- and third-generation fermions are subject to significantly stricter constraints than one would expect from a tree-level analysis without considering RGE effects. For some of them, the constraints are even stronger than those on operators involving the first-generation fermions.
In the following, we provide a more detailed analysis of the results.

%%%%%%%%%%%%%%%%%%%%
\begin{table}[t]
\center
\resizebox{\textwidth}{!}{
\renewcommand{\arraystretch}{1.0}
\begin{tabular}{|c|c|c|c|c|}
\hline 
\multicolumn{5}{|c|}{ \cellcolor{Orange!20}
RGE-improved constraints on the dimensionless WCs $c_7^i$ at $\Lambda=10$ TeV
}\\
\hline
\multirow{2}{*}{Operator} 
& \multicolumn{1}{c|}{ $|c_7^i|\equiv |\Lambda^{3} C_7^i(\Lambda)|$} 
&\multirow{2}{*}{ Operator} 
& \multicolumn{2}{c|}{ $|c_7^i|\equiv |\Lambda^{3} C_7^i(\Lambda)|$}
\\
\cline{2-2}
\cline{4-5}
& \multicolumn{1}{c|}{\cellcolor{Green!25} Same in both bases  }& 
& \cellcolor{blue!25} Up-quark flavor basis
& \cellcolor{red!25}  Down-quark flavor basis
\\
\hline
$\calO_{LH}^{11}$& {\cellcolor{gray!80}$1.92 \times 10^{-8}$}
 & $\calO_{\overline{Q}uLLH}^{1111}$ & {\cellcolor{gray!50}$2.85 \times 10^{-5}$} 
 & {\cellcolor{gray!50}$2.78 \times 10^{-5}$} 
\\
$\calO_{LeDH}^{11}$&\cellcolor{gray!40}$1.88 \times 10^{-4}$ 
& $\calO_{\overline{Q}uLLH}^{1211}$ & $8.58 \times 10^{5}$ 
&\cellcolor{gray!40} $4.48 \times 10^{-4}$ 
\\
$\calO_{DLDH1}^{11}$& {\cellcolor{gray!60}$1.07 \times 10^{-6}$} 
& $\calO_{\overline{Q}uLLH}^{1311}$ & $3.51$ 
&\cellcolor{gray!60} $6.50 \times 10^{-6}$ 
\\
$\calO_{DLDH2}^{11}$& {\cellcolor{gray!70}$9.99 \times 10^{-7}$} 
& $\calO_{\overline{Q}uLLH}^{2111}$ &\cellcolor{gray!40} $1.24 \times 10^{-4}$ 
&\cellcolor{gray!10} $2.65 \times 10^{-1}$ 
\\
$\calO_{LHW}^{11}$& {\cellcolor{gray!70}$4.26 \times 10^{-7}$} & $\calO_{\overline{Q}uLLH}^{2211}$ &\cellcolor{gray!40} $1.00 \times 10^{-4}$ 
&\cellcolor{gray!40} $1.03 \times 10^{-4}$ 
\\
\cline{1-2}
$\calO_{\overline{e}LLLH}^{{\tt (M),}2112}$ &\cellcolor{gray!30} $1.76 \times 10^{-3}$ 
& $\calO_{\overline{Q}uLLH}^{2311}$ &\cellcolor{gray!10} $3.01 \times 10^{-1}$ 
& \cellcolor{gray!60} $1.43 \times 10^{-6}$ 
\\
$\calO_{\overline{e}LLLH}^{{\tt (M),}3113}$ &\cellcolor{gray!40} $1.05 \times 10^{-4}$ 
& $\calO_{\overline{Q}uLLH}^{3111}$ &\cellcolor{gray!30} $3.24 \times 10^{-3}$ 
&\cellcolor{gray!20} $5.38 \times 10^{-2}$
\\
$\calO_{\overline{e}LLLH}^{{\tt (S),}1111}$ &\cellcolor{gray!10} $2.59 \times 10^{-1}$ 
& $\calO_{\overline{Q}uLLH}^{3211}$ & $5.65 \times 10^{1}$ 
&\cellcolor{gray!30} $2.38 \times 10^{-3}$ 
\\
$\calO_{\overline{e}LLLH}^{{\tt (S),}2112}$ &\cellcolor{gray!30} $1.25 \times 10^{-3}$ 
&$\calO_{\overline{Q}uLLH}^{3311}$ & {\cellcolor{gray!80}$5.90 \times 10^{-8}$} 
& {\cellcolor{gray!80}$5.90 \times 10^{-8}$}
\\
$\calO_{\overline{e}LLLH}^{{\tt(S),}3113}$&{\cellcolor{gray!50}$7.44 \times 10^{-5}$} & & &
\\%%%%%
\hline
$\calO_{\overline{d}uLDL}^{1111}$ &\cellcolor{gray!30} $1.92 \times 10^{-3}$ 
&$\calO_{\overline{d}LQLH1}^{1111}$ & {\cellcolor{gray!50}$6.22 \times 10^{-5}$} 
& {\cellcolor{gray!50}$6.38 \times 10^{-5}$} 
\\
$\calO_{\overline{d}uLDL}^{1211}$ & $3.58 \times 10^{2}$  
&$\calO_{\overline{d}LQLH1}^{1121}$ &\cellcolor{gray!10} $2.17 \times 10^{-1}$ 
&\cellcolor{gray!40} $2.77 \times 10^{-4}$ 
\\
$\calO_{\overline{d}uLDL}^{1311}$ & $1.54 \times 10^{1}$ 
&  $\calO_{\overline{d}LQLH1}^{1131}$ & $4.69$
&\cellcolor{gray!20} $1.72\times10^{-2}$ 
\\
$\calO_{\overline{d}uLDL}^{2111}$ & $9.21 \times 10^{3}$ 
& $\calO_{\overline{d}LQLH1}^{2111}$ &\cellcolor{gray!20} $1.06 \times 10^{-2}$ 
& $4.50 \times 10^{1}$
\\
$\calO_{\overline{d}uLDL}^{2211}$ & $4.04$ 
&  $\calO_{\overline{d}LQLH1}^{2121}$ &\cellcolor{gray!30} $2.45 \times 10^{-3}$ 
&\cellcolor{gray!30} $2.39 \times 10^{-3}$
\\
$\calO_{\overline{d}uLDL}^{2311}$ &\cellcolor{gray!10} $1.66 \times 10^{-1}$ 
& $\calO_{\overline{d}LQLH1}^{2131}$ &\cellcolor{gray!20} $5.05 \times 10^{-2}$ 
&\cellcolor{gray!10} $4.09 \times 10^{-1}$
\\
$\calO_{\overline{d}uLDL}^{3111}$ & $1.34 \times 10^{2}$ 
& $\calO_{\overline{d}LQLH1}^{3111}$ &\cellcolor{gray!20} $1.27 \times 10^{-2}$
&\cellcolor{gray!20} $3.61 \times 10^{-2}$
\\
$\calO_{\overline{d}uLDL}^{3211}$ & $1.80 $  
& $\calO_{\overline{d}LQLH1}^{3121}$ &\cellcolor{gray!30} $1.09 \times 10^{-3}$ 
&\cellcolor{gray!30} $7.91 \times 10^{-3}$
\\
$\calO_{\overline{d}uLDL}^{3311}$ &\cellcolor{gray!40} $1.33 \times 10^{-4}$  
& $\calO_{\overline{d}LQLH1}^{3131}$ &\cellcolor{gray!50}$4.04 \times 10^{-5}$
&\cellcolor{gray!50}$4.04 \times 10^{-5}$
\\%%%%%
\hline
$\calO_{\overline{d}LueH}^{1111}$ &\cellcolor{gray!20} $2.18 \times 10^{-2}$  
& $\calO_{\overline{d}LQLH2}^{1111}$ &\cellcolor{gray!40} $2.61 \times 10^{-4}$
&\cellcolor{gray!40} $2.68 \times 10^{-4}$
\\
$\calO_{\overline{d}LueH}^{1121}$ & $9.27 \times 10^{4}$  
& $\calO_{\overline{d}LQLH2}^{1121}$ & $3.57$ 
&\cellcolor{gray!30} $1.16 \times 10^{-3}$  
\\
$\calO_{\overline{d}LueH}^{1131}$ & $8.92 \times 10^{3}$  
&$\calO_{\overline{d}LQLH2}^{1131}$ & $8.56 \times 10^{1}$
&\cellcolor{gray!20} $7.20 \times 10^{-2}$ 
\\
$\calO_{\overline{d}LueH}^{2111}$ & $2.27 \times 10^{6}$  
&$\calO_{\overline{d}LQLH2}^{2111}$ &\cellcolor{gray!10} $1.75 \times 10^{-1}$
&$2.94 \times 10^{3}$
\\
$\calO_{\overline{d}LueH}^{2121}$ & $1.05 \times 10^{3}$  
&$\calO_{\overline{d}LQLH2}^{2121}$ &\cellcolor{gray!20} $4.03 \times 10^{-2}$ 
&\cellcolor{gray!20} $3.93 \times 10^{-2}$
\\
$\calO_{\overline{d}LueH}^{2131}$ & $9.58 \times 10^{1}$  
&$\calO_{\overline{d}LQLH2}^{2131}$ &\cellcolor{gray!10} $9.16 \times 10^{-1}$
&$2.66 \times 10^{1}$ 
\\
$\calO_{\overline{d}LueH}^{3111}$ & $2.72 \times 10^{6}$
&$\calO_{\overline{d}LQLH2}^{3111}$ &\cellcolor{gray!10} $2.09 \times 10^{-1}$ 
&$2.35$ 
\\
$\calO_{\overline{d}LueH}^{3121}$ & $4.67 \times 10^{2}$  
&$\calO_{\overline{d}LQLH2}^{3121}$ &\cellcolor{gray!20} $1.80 \times 10^{-2}$
&\cellcolor{gray!10} $5.14 \times 10^{-1}$
\\
$\calO_{\overline{d}LueH}^{3131}$ &\cellcolor{gray!20} $7.67 \times 10^{-2}$ 
&$\calO_{\overline{d}LQLH2}^{3131}$ &\cellcolor{gray!40} $7.33 \times 10^{-4}$ 
&\cellcolor{gray!40} $7.32 \times 10^{-4}$ 
\\
\hline
\end{tabular}
}
\caption{Limits on the dimensionless WCs of relevant $\Delta L=2$ dim-7 SMEFT operators with all possible generation combinations at  $\Lambda = 10$ TeV from $0\nu\beta\beta$ decay, after accounting for the full RGE effects.
The progressively darker gray shading indicates increasingly stronger constraints. 
}
\label{tab:limit_dim7_Ops}
\end{table}  
%%%%%%%%%%%%%%%%%%%%

For operators that do not involve a left-handed quark doublet $Q$ 
($\calO_{LH}$, 
$\calO_{LeDH}$,
$\calO_{DLDH1,2}$,
$\calO_{LHW}$, 
$\calO_{\overline{e}LLLH}^{\tt (M,S)}$, 
$\calO_{\overline{d}uLDL}$, 
and $\calO_{\overline{d}LueH}$), 
the constraints on their WCs derived from $0\nu\beta\beta$ decay are the same in both quark flavor bases. However, for operators involving a $Q$  (namely, $\calO_{\overline{Q}uLLH}$ and $\calO_{\overline{d}LQLH1,2}$), the constraints strongly depend on the choice of the quark flavor basis, except in the case where the quark fields in an operator belong to the same generation. Both features can be understood from the structure of the RGEs and their mixing behavior with the 11 first-generation-fermion operators that directly contribute to $0\nu\beta\beta$, as discussed in \cref{sec:1_gene_ope}. For instance, the constraints on the operators $\calO_{\overline{d}LQLH1}^{p1r1}$ and $\calO_{\overline{Q}uLLH}^{pr11}$ with $p,r\neq1$ primarily arise from their mixing with the neutrino mass operators via both a single and a triple product of quark Yukawa matrices ($Y_{i}$ and $Y_{i}Y_{i}Y_{i}$ with $i=u,d$) \cite{Liao:2019tep}, both of which depend on the choice of quark flavor basis. For those operators with the right-handed quarks in the first generation, i.e., 
$\mathcal{O}_{\overline{d}LQLH1}^{11w1}$ and $\mathcal{O}_{\overline{Q}uLLH}^{w111}$ with $w = 2, 3$, 
the basis dependence of the constraints arises through the non-diagonal CKM matrix elements in both the RGEs and the matching conditions. For the operators $\calO_{\overline{d}uLDL}^{pr11}$ and $\calO_{\overline{d}LueH}^{p1r1}$ with $p,r\neq 1$, their constraints mainly originate from RGE mixing with $\calO_{DLDH1}^{11}$ and $\calO_{LeDH}^{11}$, respectively, via the quark Yukawa combination $Y_u^{\dagger}Y_d$ \cite{Liao:2019tep}. However, $Y_u^{\dagger}Y_d$ remains the same in both quark bases, which makes the constraints on these two operators basis-independent. 

By incorporating the full RGE running and mixing of dim-5 and dim-7 SMEFT interactions, we find that $0\nu\beta\beta$ decay experiments can impose stringent constraints on a much broader generation combinations of dim-7 SMEFT operators than a simple tree-level analysis suggests: 55 WCs compared to 10 are constrained. Notably, for operators involving the third-generation quarks, the constraints are comparable to or even stronger than those on the corresponding first-generation-fermion operators contributing to $0\nu\beta\beta$ at the tree-level; these are highlighted in gray in \cref{tab:limit_dim7_Ops}.  
For the purely leptonic operators $\calO_{\overline{e}LLLH}^{\tt (M)}$ and $ \calO_{\overline{e}LLLH}^{\tt(S)}$,
previous studies~\cite{Cirigliano:2017djv, Liao:2019tep,Scholer:2023bnn} omitted their contributions to $0\nu\beta\beta$ decay, but we find they also receive very stringent constraints due to the running and mixing effects.  
These results highlight the importance of accurately solving and incorporating the full RGE effects of dim-5 and dim-7 SMEFT interactions  when conducting low-energy phenomenological studies of other physical processes to which these interactions may contribute.

%%%%%%%%%%%%%%%%%%%%%%%%%%%%%%%%%%%%%%
\section{Summary}
\label{sec:summary}
%%%%%%%%%%%%%%%%%%%%%%%%%%%%%%%%%%%%%%

Dimension-7 SMEFT interactions can induce processes violating lepton number by two units or baryon number by one unit, making them of significant phenomenological interest, especially for LNV processes that are otherwise suppressed by the dim-5 neutrino mass operator. Given the growing importance of RGE effects in precision phenomenological analyses, we have presented \texttt{D7RGESolver}, an automatic {\it Python}-based tool for numerically solving the full RGEs of dim-5 and dim-7 SMEFT operators induced by SM interactions. With \texttt{D7RGESolver}, the WCs of dim-5 and dim-7 operators can be rapidly and precisely evolved between any two scales above the electroweak scale. The tool supports two commonly used quark flavor bases, up-quark and down-quark flavor bases, facilitating straightforward connections with specific UV models.
We anticipate that \texttt{D7RGESolver} will serve as a valuable resource for precision studies of lepton and/or baryon number violating processes at low energies, enabling systematic investigations of new physics scenarios beyond the Weinberg operator framework.

To illustrate the importance of a complete RGE evolution in phenomenological analyses, we have taken the $0\nu\beta\beta$ decay as a concrete example. By interfacing \texttt{D7RGESolver} with \texttt{$\nu$DoBe}, we have performed a comprehensive study of $0\nu\beta\beta$ decay within the dim-5 and dim-7 SMEFT framework, incorporating the full one-loop RGE effects. Several key features emerge from the RGE-improved results. First, dim-7 operators that mix into the neutrino mass operators $\calO_{LH(5)}^{11}$, such as $\calO_{DLDH1}^{11}$, $\calO_{DLDH2}^{11}$, and $\calO_{LHW}^{11}$, receive significantly stronger constraints than those derived from their direct tree-level contributions. Additionally, operators involving second- and/or third-generation fermion fields can contribute to $0\nu\beta\beta$ through RGE mixing and consequently receive meaningful constraints, even if their tree-level contributions are absent or suppressed. Notably, for operators involving third-generation quarks or leptons, such as $\calO_{\overline{Q}uLLH}^{3311}$, $\calO_{\overline{d}LQLH1}^{3131}$, and $\calO_{\overline{e}LLLH}^{{\tt( M,S)}3113}$, the large Yukawa couplings in RGEs result in constraints on the corresponding dimensionless WCs as stringent as $10^{-8}-10^{-5}$ at a NP scale of 10 TeV. These results clearly demonstrate that including the full RGE effects leads to significant deviations from naive tree-level analyses, highlighting the necessity of a complete RGE treatment for accurate phenomenological predictions.

\vspace{2mm}		
{\bf Notes added}: During the finalization of this manuscript, a preprint \cite{Graf:2025cfk} appeared that also developed the similar idea of the RGE effect of the dim-7 operators on $0\nu\beta\beta$ decay. 

%%%%%%%%%%%%%%%%%%%%%%%%%%
\acknowledgments
%%%%%%%%%%%%%%%%%%%%%%%%%%
This work was supported 
by the Grants 
No.\,NSFC-12035008, 
No.\,NSFC-12247151, 
and No.\,NSFC-12305110.

%%%%%%%%%%%%%%%%%%%%%%%%%%
\appendix
%%%%%%%%%%%%%%%%%%%%%%%%%%

%%%%%%%%%%%%%%%%%%%%%%%%%%
\section{Summary of the RGEs for all dim-5 and dim-7 operators}
\label{sec:full_RGEs}
%%%%%%%%%%%%%%%%%%%%%%%%%%
In this Appendix, we summarize the full RGEs for dim-5 and dim-7 SMEFT operators at one-loop level, as calculated in Refs.~\cite{Liao:2016hru,Liao:2019tep,Zhang:2023kvw,Zhang:2023ndw}. The definitions of the operators appearing in these RGEs are provided in \cref{tab:dim7_SMEFT_OP}. Additionally, we also present the RGEs for the SM parameters at one-loop level \cite{Luo:2002ey}.
These RGEs are implemented in the file \texttt{beta\_function.py} within the code \texttt{D7RGESolver}. In our code, we also include the two-loop RGEs for the SM parameters~\cite{Machacek:1983tz,Machacek:1983fi,Machacek:1984zw,Luo:2002ey}. Users can choose flexibly the SM loop orders when solving the RGEs.

To display the RGEs for the WCs  briefly, 
we adopt the notation  
$ \dot{C_i} \equiv 16\pi^2{{\rm d}C}/{{\rm d}\ln\mu} $ 
and the abbreviations  
$ W_H \equiv {\rm Tr}(Y_l^{\dagger} Y_l + 3Y_u^{\dagger}Y_u  + 3Y_d^{\dagger}Y_d )$ 
and $ T_H \equiv {\rm Tr}\big[( Y_l Y^\dagger_l )^2 + 3 ( Y_{u} Y^\dagger_{u})^2 + 3 ( Y_{d} Y^\dagger_{d})^2 \big]$.
\vspace{2mm}

\noindent $\bullet$ {\textbf{SM parameters}}
\begin{subequations}
	\begin{align}
		16\pi^2\mu \frac{{\rm d} g^{\prime}}{{\rm d} \mu} 
		&= \frac{41}{6}g^{\prime3}\;,
		\\
		16\pi^2\mu \frac{{\rm d} g}{{\rm d} \mu} 
		& = - \frac{19}{6} g^3 \;,
		\\
		16\pi^2\mu \frac{{\rm d} g_s}{{\rm d} \mu} 
		& = - 7 g^3_s \;,
		\\
		16\pi^2\mu \frac{{\rm d} \mu_h^2}{{\rm d} \mu} 
		& = \mu_h^2 \Big( -\frac{3}{2} g^{\prime2} - \frac{9}{2} g^2 + 12 \lambda + 2 W_H \Big)\;,
		\\
		16\pi^2\mu \frac{{\rm d} \lambda}{{\rm d} \mu} 
		& = \frac{3}{8} g^{\prime4} + \frac{9}{8} g^4 + \frac{3}{4} g^{\prime2} g^2 - 2 T_H + ( -3g^{\prime2} - 9g^2 + 24\lambda + 4W_H ) \lambda  \;,
		\\
        \label{eq:Ylrun}
		16\pi^2\mu \frac{{\rm d} Y_l}{{\rm d} \mu} 
		& = \Big( -\frac{15}{4} g^{\prime2} - \frac{9}{4} g^2 + W_H + \frac{3}{2} Y_l Y^\dagger_l \Big)Y_l\;,
		\\
        \label{eq:Yurun}
		16\pi^2\mu \frac{{\rm d} Y_{u}}{{\rm d} \mu} 
		& = \Big[-\frac{17}{12} g^{\prime2} - \frac{9}{4}g^2 - 8g^2_s + W_H + \frac{3}{2}  ( Y_{u} Y^\dagger_{u} - Y_{d} Y^\dagger_{d} )\Big]Y_{u} \;, 
		\\
        \label{eq:Ydrun}
		16\pi^2\mu \frac{{\rm d} Y_{d}}{{\rm d} \mu} 
		& = \Big[ - \frac{5}{12} g^{\prime2} - \frac{9}{4}g^2 - 8g^2_s + W_H - \frac{3}{2} ( Y_{u} Y^\dagger_{u} - Y_{d} Y^\dagger_{d} )\Big]Y_{d}\;.  
	\end{align}
\end{subequations}

 %dim5
\noindent$\bullet~\bm{\psi^2H^2}$
\begin{align}
	\label{eq:RGE_Weinberg_dim5}
	\dot{C}_{LH5}^{pr} & =  
	\frac{1}{2}(-3g^2+4\lambda+2W_H)C_{LH5}^{pr}
	-\frac{3}{2}[C_{LH5} Y_lY_l^\dagger]^{pr}
	+ \mu_h^2 \big\{8 C_{LH}^{pr} +2[C_{LeDH}Y_l^{\dagger}]^{pr} 
	\nonumber
	\\
	& +{\frac{3}{2} g^2\big(2 C_{DLDH1}^{ pr}+C_{DLDH2}^{pr}\big)}
	+[C_{DLDH1}Y_l Y_l^{\dagger}]^{pr}
	-\frac{1}{2} [C_{DLDH2}Y_l Y_l^{\dagger}]^{pr} 
	\nonumber
	\\
	& -[Y_l]_{ts}\big(3 C_{\overline{e} LLL H}^{{\tt (S),}st pr}+2C_{\overline{e} LLL H}^{{\tt (M),}st pr}\big) 
	-3[Y_d]_{ts} C_{\overline{d}L Q L H 1}^{ s p t r} 
	+{6 [Y_u^{\dagger}]_{st} C_{\overline{Q} u LL H}^{ ts pr}}\big\} 
	+ p \leftrightarrow r \;.
\end{align}
Note that the definition of the dim-5 Weinberg operator in this work differs from that in Ref.~\cite{Zhang:2023kvw} by a Hermitian conjugation. Consequently, its WC ($C_{LH5}$) in this study is related to the one in that paper by a complex conjugation.

%dim7: OLH
\noindent$\bullet~\bm{\psi^2H^4}$
\begin{align}
	% \label{eq:psi2H4}
	\befun^{pr}_{LH} & =
	- \frac{1}{4}(3g^{\prime2} + 15g^2 -80\lambda-8 W_H ) C^{pr}_{L H} 
	- \frac{3}{2} [C_{LH} Y_l Y^\dagger_l]^{pr} 
	+ \big( 2\lambda - \frac{3}{2} g^2 \big) 
	[C_{LeDH} Y^\dagger_l]^{pr} 
	\nonumber
	\\
	& + [C_{L e DH} Y^\dagger_l Y_l Y^\dagger_l]^{pr} 
	- \frac{3}{4} g^2 (g^2 - 4 \lambda) C^{pr}_{DLDH1} + \lambda [C_{DLDH1} Y_l Y^\dagger_l]^{pr} 
	\nonumber
	\\
	& -[C_{DLDH1} Y_l Y^\dagger_l Y_lY^\dagger_l]^{pr}
	- \frac{3}{8} ( g^{\prime4} + 2 g^{\prime2} g^2 + 3g^4 - 4g^2 \lambda) C^{pr}_{DLDH2} 
	\nonumber
	\\
	&-\frac{1}{2}\lambda[C_{DLDH2}Y_l Y^\dagger_l]^{pr} 
    -[C_{DLDH2} Y_l Y^\dagger_l Y_l Y^\dagger_l]^{pr}
	- 3 g^3 C^{pr}_{LHW} 
	- 6 g [C_{LHW} Y_l Y^\dagger_l]^{pr} 
	\nonumber
	\\
	& - 3 C^{{\tt (S),}stpr}_{\overline{e}LLLH} 
	[\lambda Y_l -Y_l Y^\dagger_l Y_l]_{ts} 
    - 2C^{{\tt (M),}stpr}_{\overline{e}LLLH} 
	[\lambda Y_l - Y_l Y^\dagger_l Y_l]_{ts}  
	\nonumber
	\\
	& - 3 C^{sptr}_{\overline{d}LQLH1} 
	[\lambda Y_d -Y_d Y^\dagger_d Y_d]_{ts}
    + 6 C^{stpr}_{\overline{Q}uLL H} 
	[\lambda Y^\dagger_u - Y^\dagger_u Y_u Y^\dagger_u]_{ts} 
	+ p \leftrightarrow r \;.
	\label{eq:RGE_Weinberg_dim7}
\end{align}
%dim7: OLeDH
\noindent$\bullet~\bm{\psi^2H^3D}$
\begin{align}
	\befun^{pr}_{L e DH} &= 
	- \frac{3}{2} (3g^{\prime2} - 4\lambda -2W_H) C^{pr}_{LeDH} 
	+ [Y^{\tt T}_l C_{L e DH} Y^\dagger_l]^{rp} 
	+ 4 [C_{L e DH} Y^\dagger_l Y_l]^{pr} 
	+ \frac{1}{2} [C^{\tt T}_{LeDH} Y_l Y^\dagger_l]^{rp} 
	\nonumber
	\\
	& + ( 3g^{\prime2} - g^2) [C_{DLDH1} Y_l]^{pr} 
	- 2 [C_{DLDH1} Y_l Y^\dagger_l Y_l]^{pr} 
	+ \frac{1}{8}(7g^{\prime2} - 17g^2 -8 \lambda)[C_{DLDH2}Y_l]^{pr} 
	\nonumber
	\\
	& - 2 [C_{DLDH2}Y_l Y^\dagger_l Y_l]^{pr}  
	- \frac{1}{2}[Y^{\tt T}_l C_{DLDH2} Y_l Y^\dagger_l]^{rp} 
	- 6 C^{sptr}_{\overline{d}LueH} [Y^\dagger_{u} Y_{d}]_{ts} \;.
\end{align}
%dim7:ODLDH1,2
\noindent$\bullet~\bm{\psi^2H^2D^2}$
\begin{align}
	\befun^{pr}_{DLDH1} &=
	-\frac{1}{4} (3g^{\prime2} - 11g^2 - 4W_H) C^{pr}_{DLDH1} 
	+\frac{7}{2} [C_{DLDH1} Y_l Y^\dagger_l]^{pr} 
	\nonumber
	\\
	&-\frac{1}{8} (11 g^{\prime2} + 11g^2 +8\lambda) C^{pr}_{DLDH2} 
	+6C^{stpr}_{\overline{d}uLDL}[Y^\dagger_u Y_d]_{ts} 
	+ p \leftrightarrow r \;,
	\\
	\befun^{pr}_{DLDH2} & = 
	-4g^2 C^{pr}_{DLDH1} 
	- 4 [C_{DLDH1} Y_l Y^\dagger_l]^{pr} 
	+\frac{1}{2} (4 g^{\prime2} + g^2 + 4\lambda + 2W_H) C^{pr}_{DLDH2} 
	\nonumber
	\\
	& -\frac{3}{2} [C_{DLDH2} Y_l Y^\dagger_l]^{pr} 
	+ p \leftrightarrow r \;.
\end{align}
%dim7:OLHW,B
\noindent$\bullet~\bm{\psi^2H^2X}$
\begin{align}
	\befun^{pr}_{LHW} &=
	\frac{1}{2} g^3 C^{pr}_{DLDH1} 
	- \frac{1}{4} g[C_{DLDH1} Y_l Y^\dagger_l]^{rp}
	+ \frac{1}{2} g[C_{DLDH1} Y_l Y^\dagger_l]^{pr} 
	+ \frac{5}{8} g^3 C^{pr}_{DLDH2} 
	\nonumber
	\\
	& +\frac{3}{4}g[C_{DLDH2} Y_l Y^\dagger_l]^{pr} 
	+ \frac{1}{8}g[C_{DLDH2} Y_l Y^\dagger_l]^{rp} 
	- \frac{1}{2}(4g^{\prime2}-9g^2-8\lambda-4W_H)C^{pr}_{LHW} 
	\nonumber
	\\
	& + \frac{7}{2} g^2 C^{rp}_{LHW} 
	+\frac{9}{2}[C_{LHW} Y_l Y^\dagger_l]^{pr}  
	+2[C_{LHW} Y_l Y^\dagger_l]^{rp} 
	-\frac{3}{2}[C^{\tt T}_{LHW}Y_l Y^\dagger_l]^{rp} 
	\nonumber
	\\
	& + 3 g^{\prime} g C^{pr}_{LHB} 
	- \frac{1}{4} g \big( 3 C^{{\tt (S),}stpr}_{\overline{e}LLLH} 
	+ C^{{\tt (A),}stpr}_{\overline{e}LLLH} 
	- 2C^{{\tt (M),}stpr}_{\overline{e}LLLH}\big) 
	[Y_l]_{ts} 
	\nonumber
	\\
	& - \frac{3}{4} g C^{srtp}_{\overline{d}LQLH1} [Y_{d}]_{ts} 
	- \frac{3}{4} g \big( C^{sptr}_{\overline{d}LQLH2} + C^{srtp}_{\overline{d}LQLH2} \big) 
	[Y_d]_{ts} \;,
	\\
	\befun^{pr}_{LHB} &= 
	- \frac{1}{8} g^{\prime}[C_{DLDH1} Y_l Y^\dagger_l]^{pr}
	- \frac{3}{16} g^{\prime} [C_{DLDH2} Y_l Y^\dagger_l]^{pr}
	+ 3 g^{\prime} g C^{pr}_{LHW} 
	- \frac{3}{2} [C_{LHB} Y_l Y^\dagger_l]^{pr} 
	\nonumber
	\\
	& + \frac{1}{12} ( 47g^{\prime2} - 30g^2 + 24 \lambda + 12W_H ) C^{pr}_{LHB}
	+ \frac{3}{8} g^{\prime} \big( 
	C^{{\tt (A),}stpr}_{\overline{e}LLLH} 
	- 2 C^{{\tt (M),}stpr}_{\overline{e} LLLH} \big) [Y_l]_{ts}
	\nonumber
	\\
	& - \frac{1}{8} g^{\prime} C^{sptr}_{\overline{d} LQLH1}[Y_{d}]_{ts} 
	- p \leftrightarrow r \;.
\end{align}
%dim7: OeLLLH
%\XD{stophere.}
\noindent$\bullet~\bm{\psi^4H}$
\begin{align}
	\label{eq:psi4H}
	\befun^{{\tt (S),}prst}_{\overline{e}LLLH} &= ( g^{\prime2} - g^2 ) \big\{ [ Y^\dagger_l ]_{pr} C^{st}_{DLDH1} +  [ Y^\dagger_l ]_{ps} C^{rt}_{DLDH1} +  [ Y^\dagger_l ]_{pt} C^{rs}_{DLDH1} \big\}
	\nonumber
	\\
	& - \frac{1}{2} \big\{ [C_{DLDH1} Y_l Y^\dagger_l ]^{tr} [ Y^\dagger_l ]_{ps} + [C_{DLDH1} Y_l Y^\dagger_l ]^{ts} [ Y^\dagger_l ]_{pr} + [C_{DLDH1} Y_l Y^\dagger_l ]^{sr} [ Y^\dagger_l ]_{pt} 
	\nonumber
	\\
	& + [C_{DLDH1} Y_l Y^\dagger_l ]^{st} [ Y^\dagger_l ]_{pr} + [C_{DLDH1} Y_l Y^\dagger_l ]^{rs} [ Y^\dagger_l ]_{pt} + [C_{DLDH1} Y_l Y^\dagger_l ]^{rt} [ Y^\dagger_l ]_{ps} \big\}
	\nonumber
	\\
	&+  \frac{1}{2} ( g^{\prime2} - 2g^2 )  \big\{  [ Y^\dagger_l ]_{pr} C^{st}_{DLDH2} +  [ Y^\dagger_l ]_{ps} C^{rt}_{DLDH2} +  [ Y^\dagger_l ]_{pt} C^{rs}_{DLDH2} \big\} 
	\nonumber
	\\
	& - \frac{1}{4} \big\{ [C_{DLDH2} Y_l Y^\dagger_l ]^{tr} [ Y^\dagger_l ]_{ps} + [C_{DLDH2} Y_l Y^\dagger_l ]^{ts} [ Y^\dagger_l ]_{pr} + [C_{DLDH2} Y_l Y^\dagger_l ]^{sr} [ Y^\dagger_l ]_{pt} 
	\nonumber
	\\
	& + [C_{DLDH2} Y_l Y^\dagger_l ]^{st} [ Y^\dagger_l ]_{pr} + [C_{DLDH2} Y_l Y^\dagger_l ]^{rs} [ Y^\dagger_l ]_{pt} + [C_{DLDH2} Y_l Y^\dagger_l ]^{rt} [ Y^\dagger_l ]_{ps} \big\} 
	\nonumber
	\\
	& - 2g \big\{ [ Y^\dagger_l ]_{pr} \big( C^{st}_{LHW} +  C^{ts}_{LHW} \big) + [ Y^\dagger_l ]_{ps} \big( C^{rt}_{LHW} + C^{tr}_{LHW} \big)  
	\nonumber
	\\
	& + [ Y^\dagger_l ]_{pt} \big( C^{rs}_{LHW} + C^{sr}_{LHW} \big) \big\} - \frac{1}{4} ( 9g^{\prime2} - 9g^2 - 4W_H ) C^{{\tt (S),}prst}_{\overline{e}LLLH} + 3 [ Y^\dagger_l Y_l ]_{pv} C^{{\tt (S),}vrst}_{\overline{e}LLLH} 
	\nonumber
	\\
	& + [ Y_l ]_{vw} \big\{ C^{{\tt (S),}vwst}_{\overline{e}LLLH} [ Y^\dagger_l ]_{pr} + C^{{\tt (S),}vwtr}_{\overline{e}LLLH} [ Y^\dagger_l ]_{ps} + C^{{\tt (S),}vwrs}_{\overline{e}LLLH} [ Y^\dagger_l ]_{pt} \big\}
	\nonumber
	\\
	& + \frac{1}{2} [ Y_l ]_{vw} \big\{ C^{{\tt (S),}pvst}_{\overline{e}LLLH} [ Y^\dagger_l ]_{wr} + C^{{\tt (S),}prvt}_{\overline{e}LLLH} [ Y^\dagger_l ]_{ws} + C^{{\tt (S),}prsv}_{\overline{e}LLLH} [ Y^\dagger_l ]_{wt} \big\}
	\nonumber
	\\
	& + \frac{2}{3} \big\{ [ Y_l Y^\dagger_l  ]_{vr} \big( C^{{\tt (M),}pvst}_{\overline{e}LLLH} + C^{{\tt (M),}pvts}_{\overline{e}LLLH} \big) +  [ Y_l Y^\dagger_l  ]_{vs} \big( 2C^{{\tt (M),}pvtr}_{\overline{e}LLLH} + C^{{\tt (M),}prvt}_{\overline{e}LLLH} \big) 
	\nonumber
	\\
	&+ [ Y_l Y^\dagger_l  ]_{vt} \big( 2 C^{{\tt (M),}pvsr}_{\overline{e}LLLH}  +  C^{{\tt (M),}prvs}_{\overline{e}LLLH} \big) \big\} + \frac{1}{3} [ Y_l ]_{vw} \big\{ \big( C^{{\tt (M),}wvst}_{\overline{e}LLLH} + C^{{\tt (M),}wvts}_{\overline{e}LLLH}\big)[ Y^\dagger_l ]_{pr}  
	\nonumber
	\\
	& +  \big( C^{{\tt (M),}wvrt}_{\overline{e}LLLH} + C^{{\tt (M),}wvtr}_{\overline{e}LLLH} \big) [ Y^\dagger_l ]_{ps} +  \big( C^{{\tt (M),}wvrs}_{\overline{e}LLLH} + C^{{\tt (M),}wvsr}_{\overline{e}LLLH} \big) [ Y^\dagger_l ]_{pt}   \big\}
	\nonumber
	\\
	& + \frac{1}{2}  [Y_{d} ]_{vw}  \big\{ \big( C^{wsvt}_{\overline{d}LQLH1} + C^{wtvs}_{\overline{d}LQLH1} \big) [ Y^\dagger_l ]_{pr} + \big( C^{wrvt}_{\overline{d}LQLH1} + C^{wtvr}_{\overline{d}LQLH1} \big) [ Y^\dagger_l ]_{ps} 
	\nonumber
	\\
	& + \big( C^{wsvr}_{\overline{d}LQLH1} + C^{wrvs}_{\overline{d}LQLH1} \big) [ Y^\dagger_l ]_{pt} \big\} -  [ Y^\dagger_{u} ]_{vw} \big\{ \left( C^{wvst}_{\overline{Q}uLL H} + C^{wvts}_{\overline{Q}uLL H} \right) [ Y^\dagger_l ]_{pr}  
	\nonumber
	\\
	&+ \big( C^{wvrt}_{\overline{Q}uLLH} +  C^{wvtr}_{\overline{Q}uLL H} \big) 
    [ Y^\dagger_l ]_{ps} 
    + \big( C^{wvrs}_{\overline{Q}uLL H} + C^{wvsr}_{\overline{Q}uLLH}\big) [ Y^\dagger_l ]_{pt} \big\} \;,
	\\%%%
	\befun^{{\tt (A),}prst}_{\overline{e}LLLH} &= \frac{1}{6} \big\{  [ C_{DLDH1} Y_l Y^\dagger_l ]^{ts} [ Y^\dagger_l ]_{pr} -  [ C_{DLDH1} Y_l Y^\dagger_l ]^{tr} [ Y^\dagger_l ]_{ps} + [ C_{DLDH1} Y_l Y^\dagger_l ]^{sr} [ Y^\dagger_l ]_{pt} 
	\nonumber
	\\
	& - [ C_{DLDH1} Y_l Y^\dagger_l ]^{st} [ Y^\dagger_l ]_{pr} + [ C_{DLDH1} Y_l Y^\dagger_l ]^{rt} [ Y^\dagger_l ]_{ps} - [ C_{DLDH1} Y_l Y^\dagger_l ]^{rs} [ Y^\dagger_l ]_{pt} \big\}
	\nonumber
	\\
	& - \frac{5}{12} \big\{  [ C_{DLDH2} Y_l Y^\dagger_l ]^{ts} [ Y^\dagger_l ]_{pr} -  [ C_{DLDH2} Y_l Y^\dagger_l ]^{tr} [ Y^\dagger_l ]_{ps} + [ C_{DLDH2} Y_l Y^\dagger_l ]^{sr} [ Y^\dagger_l ]_{pt} 
	\nonumber
	\\
	& - [ C_{DLDH2} Y_l Y^\dagger_l ]^{st} [ Y^\dagger_l ]_{pr} + [ C_{DLDH2} Y_l Y^\dagger_l ]^{rt} [ Y^\dagger_l ]_{ps} - [ C_{DLDH2} Y_l Y^\dagger_l ]^{rs} [ Y^\dagger_l ]_{pt} \big\}
	\nonumber
	\\
	& - 4g \big\{  [ Y^\dagger_l ]_{pr} C^{st}_{LHW}  - [ Y^\dagger_l ]_{pr} C^{ts}_{LHW} - [ Y^\dagger_l ]_{ps} C^{rt}_{LHW} + [ Y^\dagger_l ]_{ps} C^{tr}_{LHW} 
	\nonumber
	\\
	& + [ Y^\dagger_l ]_{pt} C^{rs}_{LHW} - [ Y^\dagger_l ]_{pt} C^{sr}_{LHW}   \big\} + 12g^{\prime} \big\{ [ Y^\dagger_l ]_{pr} C^{st}_{LHB} + [ Y^\dagger_l ]_{ps} C^{tr}_{LHB}  
	\nonumber
	\\
	& + [ Y^\dagger_l ]_{pt} C^{rs}_{LHB}  \big\} - \frac{1}{4} \left( 9g^{\prime2} + 39g^2 - 4W_H \right) C^{{\tt (A),}prst}_{\overline{e}LLLH} +  [ Y_l ]_{vw} \big\{ [ Y^\dagger_l ]_{pr} C^{{\tt (A),}wvst}_{\overline{e}LLLH} 
	\nonumber
	\\
	& -[ Y^\dagger_l ]_{ps} C^{{\tt (A),}wvrt}_{\overline{e}LLLH} + [ Y^\dagger_l ]_{pt} C^{{\tt (A),}wvrs}_{\overline{e}LLLH}  \big\}  + \frac{1}{2} [ Y_l ]_{vw} \big\{ C^{{\tt (A),}pvst}_{\overline{e}LLLH} [ Y^\dagger_l ]_{wr} 
	\nonumber
	\\
	& +  C^{{\tt (A),}prvt}_{\overline{e}LLLH} [ Y^\dagger_l ]_{ws} + C^{{\tt (A),}prsv}_{\overline{e}LLLH} [ Y^\dagger_l ]_{wt} \big\}
	+ 3 [ Y^\dagger_l Y_l ]_{pv} C^{{\tt (A),}vrst}_{\overline{e}LLLH} -  2[ Y_l Y^\dagger_l  ]_{vr} C^{{\tt (M),}pvst}_{\overline{e}LLLH} 
	\nonumber
	\\
	& + 2 [ Y_l Y^\dagger_l  ]_{vr} C^{{\tt (M),}pvts}_{\overline{e}LLLH} + 2 [ Y_l Y^\dagger_l  ]_{vs} C^{{\tt (M),}prvt}_{\overline{e}LLLH} -  2[ Y_l Y^\dagger_l  ]_{vt} C^{{\tt (M),}prvs}_{\overline{e}LLLH} 
	\nonumber
	\\
	& + [ Y_l]_{vw} \big\{ [ Y^\dagger_l]_{pr} \big( C^{{\tt (M),}wvst}_{\overline{e}LLLH} - C^{{\tt (M),}wvts}_{\overline{e}LLLH} \big) - [ Y^\dagger_l ]_{ps} \big( C^{{\tt (M),}wvrt}_{\overline{e}LLLH} - C^{{\tt (M),}wvtr}_{\overline{e}LLLH} \big) 
	\nonumber
	\\
	& +   [ Y^\dagger_l ]_{pt} \big( C^{{\tt (M),}wvrs}_{\overline{e}LLLH} - C^{{\tt (M),}wvsr}_{\overline{e}LLLH} \big)  \big\}  + \frac{1}{2} [ Y_{d} ]_{vw} \big\{ \big( C^{wsvt}_{\overline{d}LQLH1} - C^{wtvs}_{\overline{d}LQLH1} \big) [ Y^\dagger_l ]_{pr} 
	\nonumber
	\\
	& - \big( C^{wrvt}_{\overline{d}LQLH1} - C^{wtvr}_{\overline{d}LQLH1} \big) [ Y^\dagger_l ]_{ps} + \big(  C^{wrvs}_{\overline{d}LQLH1} - C^{wsvr}_{\overline{d}LQLH1} \big) [Y^\dagger_l ]_{pt} \big\}
	\nonumber
	\\
	& + [ Y_{d} ]_{vw} \big\{ \big( C^{wsvt}_{\overline{d}L QLH2} - C^{wtvs}_{\overline{d}L QLH2} \big) [ Y^\dagger_l ]_{pr} - \big( C^{wrvt}_{\overline{d}L QLH2} - C^{wtvr}_{\overline{d}L QLH2} \big) [ Y^\dagger_l ]_{ps} 
	\nonumber
	\\
	& +  \big(  C^{wrvs}_{\overline{d}L QLH2} - C^{wsvr}_{\overline{d}L QLH2} \big) [ Y^\dagger_l ]_{pt} \big\} - [ Y^\dagger_{u} ]_{vw} \big\{ \big( C^{wvst}_{\overline{Q}uLL H} - C^{wvts}_{\overline{Q}uLL H} \big) [ Y^\dagger_l ]_{pr}   
	\nonumber
	\\
	& -\big( C^{wvrt}_{\overline{Q}uLL H} -  C^{wvtr}_{\overline{Q}uLL H} \big) [ Y^\dagger_l ]_{ps} + \big( C^{wvrs}_{\overline{Q}uLL H} - C^{wvsr}_{\overline{Q}uLL H} \big) [ Y^\dagger_l ]_{pt} \big\} \;,
    \label{eq:CeLLLHA}
	\\%%%
	\befun^{{\tt (M),}prst}_{\overline{e}LLLH} &= - \frac{3}{2} ( g^{\prime2} + g^2 ) \big\{ [ Y^\dagger_l ]_{pr} C^{st}_{DLDH1} - [ Y^\dagger_l ]_{pt} C^{rs}_{DLDH1} \big\}  - \frac{1}{6} \big\{ 4[ C_{DLDH1} Y_l Y^\dagger_l ]^{ts} [ Y^\dagger_l ]_{pr}  
	\nonumber
	\\
	& -  [ C_{DLDH1} Y_l Y^\dagger_l ]^{tr} [ Y^\dagger_l ]_{ps} - 5 [ C_{DLDH1} Y_l Y^\dagger_l ]^{sr} [ Y^\dagger_l ]_{pt} + 5 [ C_{DLDH1} Y_l Y^\dagger_l ]^{st} [ Y^\dagger_l ]_{pr} 
	\nonumber
	\\
	& +  [ C_{DLDH1} Y_l Y^\dagger_l ]^{rt} [ Y^\dagger_l ]_{ps} - 4 [ C_{DLDH1} Y_l Y^\dagger_l ]^{rs} [ Y^\dagger_l ]_{pt} \big\} - \frac{3}{4} g^{\prime2} \big\{ [ Y^\dagger_l ]_{pr} C^{st}_{DLDH2} 
	\nonumber
	\\
	& -  [ Y^\dagger_l ]_{pt} C^{rs}_{DLDH2} \big\}  - \frac{1}{12} \big\{ 7[ C_{DLDH2} Y_l Y^\dagger_l ]^{ts} [ Y^\dagger_l ]_{pr} + 5  [ C_{DLDH2} Y_l Y^\dagger_l ]^{tr} [ Y^\dagger_l ]_{ps}   
	\nonumber
	\\
	& - 2 [ C_{DLDH2} Y_l Y^\dagger_l ]^{sr} [ Y^\dagger_l ]_{pt} + 2 [ C_{DLDH2} Y_l Y^\dagger_l ]^{st} [ Y^\dagger_l ]_{pr} - 5[ C_{DLDH2} Y_l Y^\dagger_l ]^{rt} [ Y^\dagger_l ]_{ps}
	\nonumber
	\\
	& - 7 [ C_{DLDH2} Y_l Y^\dagger_l ]^{rs} [ Y^\dagger_l ]_{pt} \big\} + g \big\{ 5 [ Y^\dagger_l ]_{pr} C^{st}_{LHW} + [ Y^\dagger_l ]_{pr} C^{ts}_{LHW} 
	\nonumber
	\\
	& +  4 [ Y^\dagger_l ]_{ps} C^{rt}_{LHW} - 4 [ Y^\dagger_l ]_{ps} C^{tr}_{LHW} -[ Y^\dagger_l ]_{pt} C^{rs}_{LHW} - 5 [ Y^\dagger_l ]_{pt} C^{sr}_{LHW} \big\}
	\nonumber
	\\
	& - 6g^{\prime} \big\{ [ Y^\dagger_l ]_{pr} C^{st}_{LHB} + 2[ Y^\dagger_l ]_{ps} C^{rt}_{LHB} + [ Y^\dagger_l ]_{pt} C^{rs}_{LHB} \big\}  + 3 \big\{ C^{{\tt (S),}pvst}_{\overline{e}LLLH} [ Y_l Y^\dagger_l ]_{vr} 
	\nonumber
	\\
	& - C^{{\tt (S),}prsv}_{\overline{e}LLLH} [ Y_l Y^\dagger_l ]_{vt} \big\}  + \frac{3}{2}[ Y_l ]_{vw} \big\{ [ Y^\dagger_l ]_{pr} C^{{\tt (S),}wvst}_{\overline{e}LLLH} - [ Y^\dagger_l ]_{pt} C^{{\tt (S),}wvrs}_{\overline{e}LLLH} \big\}
	\nonumber
	\\
	& - C^{{\tt (A),}pvst}_{\overline{e}LLLH} [ Y_l Y^\dagger_l ]_{vr} + 2 C^{{\tt (A),}prvt}_{\overline{e}LLLH} [ Y_l Y^\dagger_l ]_{vs} - C^{{\tt (A),}prsv}_{\overline{e}LLLH} [ Y_l Y^\dagger_l ]_{vt} + \frac{1}{2} [ Y_l ]_{vw} 
	\nonumber
	\\
	& \times \big\{ [ Y^\dagger_l ]_{pr} C^{{\tt (A),}wvst}_{\overline{e}LLLH} + 2[ Y^\dagger_l ]_{ps} C^{{\tt (A),}wvrt}_{\overline{e}LLLH}  + [ Y^\dagger_l ]_{pt} C^{{\tt (A),}wvrs}_{\overline{e}LLLH} \big\} + 3 [ Y^\dagger_l Y_l ]_{pv} C^{{\tt (M),}vrst}_{\overline{e}LLLH}
	\nonumber
	\\
	& - \frac{1}{4} \left( 9g^{\prime2} + 15g^2 - 4W_H \right) C^{{\tt (M),}prst}_{\overline{e}LLLH} + [ Y_l ]_{vw} \big\{ [ Y^\dagger_l ]_{pr} C^{{\tt (M),}wvst}_{\overline{e}LLLH} + [ Y^\dagger_l ]_{ps} C^{{\tt (M),}wvrt}_{\overline{e}LLLH} 
	\nonumber
	\\
	& - [ Y^\dagger_l ]_{ps} C^{{\tt (M),}wvtr}_{\overline{e}LLLH}  - [ Y^\dagger_l ]_{pt} C^{{\tt (M),}wvsr}_{\overline{e}LLLH} \big\}  + \frac{1}{2} [ Y_l ]_{vw} \big\{ C^{{\tt (M),}pvst}_{\overline{e}LLLH} [ Y^\dagger_l ]_{wr} 
	\nonumber
	\\
	& + C^{{\tt (M),}prvt}_{\overline{e}LLLH} [ Y^\dagger_l ]_{ws} + C^{{\tt (M),}prsv}_{\overline{e}LLLH} [ Y^\dagger_l ]_{wt} \big\} + \frac{1}{2} [ Y_{d} ]_{vw} \big\{ \big( 2C^{wsvt}_{\overline{d}LQLH1} + C^{wtvs}_{\overline{d}LQLH1} \big) [ Y^\dagger_l ]_{pr} 
	\nonumber
	\\
	& +  \big( C^{wrvt}_{\overline{d}LQLH1} - C^{wtvr}_{\overline{d}LQLH1} \big) [ Y^\dagger_l ]_{ps} - \big(  C^{wrvs}_{\overline{d}LQLH1} + 2 C^{wsvr}_{\overline{d}LQLH1} \big) [ Y^\dagger_l ]_{pt} \big\}
	\nonumber
	\\
	& + \frac{1}{2} [ Y_{d} ]_{vw} \big\{ \big( C^{wsvt}_{\overline{d}L QLH2} - C^{wtvs}_{\overline{d}L QLH2} \big) [ Y^\dagger_l ]_{pr} + 2 \big( C^{wrvt}_{\overline{d}L QLH2} - C^{wtvr}_{\overline{d}L QLH2} \big) [ Y^\dagger_l ]_{ps} 
	\nonumber
	\\
	& +  \big(  C^{wrvs}_{\overline{d}L QLH2} - C^{wsvr}_{\overline{d}L QLH2} \big) [ Y^\dagger_l ]_{pt} \big\} -  [ Y^\dagger_{u} ]_{vw} \big\{ \big( 2C^{wvst}_{\overline{Q}uLL H} + C^{wvts}_{\overline{Q}uLL H} \big) [ Y^\dagger_l ]_{pr}   
	\nonumber
	\\
	& +  \big( C^{wvrt}_{\overline{Q}uLL H} -  C^{vwtr}_{\overline{Q}uLL H} \big) [ Y^\dagger_l ]_{ps} - \big( C^{vwrs}_{\overline{Q}uLL H} + 2 C^{vwsr}_{\overline{Q}uLL H} \big) [ Y^\dagger_l ]_{pt} \big\} \;,
	\\
	\befun^{prst}_{\overline{d}LQLH1} &= - 3 g^2 [ Y^\dagger_{d} ]_{ps} \big( 2C^{rt}_{DLDH1} + C^{rt}_{DLDH2} \big) - 3 [ Y^\dagger_{d} ]_{ps} \big\{ [ C_{DLDH1} Y_l Y^\dagger_l ]^{rt} + [ C_{DLDH1} Y_l Y^\dagger_l ]^{tr} \big\}
	\nonumber
	\\
	& - \frac{3}{2} [ Y^\dagger_{d} ]_{ps} \big\{ [ C_{DLDH2} Y_l Y^\dagger_l ]^{rt} + [ C_{DLDH2} Y_l Y^\dagger_l ]^{tr} \big\} + 8 g [ Y^\dagger_{d} ]_{ps} [ C^{rt}_{LHW} - C^{tr}_{LHW} ] 
	\nonumber
	\\
	& - \frac{8}{3} g^{\prime} [ Y^\dagger_{d} ]_{ps} C^{rt}_{LHB} + 2[ Y^\dagger_{d} ]_{ps} [ Y_l ]_{vw} \big( 3 C^{{\tt (S),}wvrt}_{\overline{e}LLLH} + C^{{\tt (M),}wvrt}_{\overline{e}LLLH} + C^{{\tt (M),}wvtr}_{\overline{e}LLLH} \big)  
	\nonumber
	\\
	& - \frac{1}{36} ( 41g^{\prime2} + 63g^2 + 96 g^2_s - 36W_H ) C^{prst}_{\overline{d}LQLH1} + \frac{4}{9} ( 5g^{\prime2} +9g^2 -12g^2_s ) C^{ptsr}_{\overline{d}LQLH1} 
	\nonumber
	\\
	& + \frac{1}{2} C^{prvt}_{\overline{d}LQLH1}  \big\{ 5[ Y_{d} Y^\dagger_{d} ]_{vs}  + [ Y_{u} Y^\dagger_{u} ]_{vs} \big\}  + 3 C^{vrst}_{\overline{d}LQLH1} [ Y^\dagger_{d} Y_{d} ]_{pv} + \frac{1}{2} C^{pvst}_{\overline{d}LQLH1} [ Y_l Y^\dagger_l ]_{vr} 
	\nonumber
	\\
	& - \frac{3}{2} C^{prsv}_{\overline{d}LQLH1} [ Y_l Y^\dagger_l ]_{vt} + 3 [ Y^\dagger_{d} ]_{ps} [ Y_{d} ]_{vw} \big( C^{wrvt}_{\overline{d}LQLH1} + C^{wtvr}_{\overline{d}LQLH1} \big)
	\nonumber
	\\
	& + 2g^2 \big( 2 C^{prst}_{\overline{d}LQLH2} + C^{ptsr}_{\overline{d}LQLH2} \big) + 2C^{pvst}_{\overline{d}LQLH2} [ Y_l Y^\dagger_l ]_{vr} - 2C^{prsv}_{\overline{d}LQLH2} [ Y_l Y^\dagger_l ] _{vt} 
	\nonumber
	\\
	& - 2C^{prvw}_{\overline{d}LueH} [ Y^\dagger_{u} ]_{vs}  [ Y^\dagger_l ]_{wt} + 2\big( C^{vwrt}_{\overline{Q}uLL H} + C^{vwtr}_{\overline{Q}uLL H} \big) \big\{ [ Y^\dagger_{d} ]_{pv} [ Y^\dagger_{u} ]_{ws}  - 3 [ Y^\dagger_{d} ]_{ps} [ Y^\dagger_{u} ]_{wv} \big\}
	\nonumber
	\\
	& - 6 g^2 C^{pvrt}_{\overline{d}uLDL} [ Y^\dagger_{u} ]_{vs} - 2C^{pvrw}_{\overline{d}uLDL} [ Y^\dagger_{u} ]_{vs} [ Y_l Y^\dagger_l ]_{wt} - 2C^{pvtw}_{\overline{d}uLDL} [ Y^\dagger_{u} ]_{vs} [ Y_l Y^\dagger_l ]_{wr}  \;,
	\\%%%
	\befun^{prst}_{\overline{d}LQLH2} &= \frac{1}{3} ( g^{\prime2} + 9g^2 ) [ Y^\dagger_{d} ]_{ps} C^{rt}_{DLDH1} + [ Y^\dagger_{d} ]_{ps} \big\{ [ C_{DLDH1} Y_l Y^\dagger_l ]^{rt} + 2[ C_{DLDH1} Y_l Y^\dagger_l ]^{tr} \big\}
	\nonumber
	\\
	& + \frac{1}{6} g^{\prime2} [ Y^\dagger_{d} ]_{ps} C^{rt}_{DLDH2} + \frac{1}{2} [ Y^\dagger_{d} ]_{ps} \big\{ 4[ C_{DLDH2} Y_l Y^\dagger_l ]^{rt} - [ C_{DLDH2} Y_l Y^\dagger_l ]^{tr} \big\} 
	\nonumber
	\\
	& - 2g [ Y^\dagger_{d} ]_{ps} \big( 5C^{rt}_{LHW} + C^{tr}_{LHW} \big) + \frac{4}{3} g^{\prime} [ Y^\dagger_{d} ]_{ps} C^{rt}_{LHB}  - [ Y^\dagger_{d} ]_{ps} [ Y_l ]_{vw} \big( 3 C^{{\tt (S),}wvrt}_{\overline{e}LLLH} 
	\nonumber
	\\
	& -  3C^{{\tt (A),}wvrt}_{\overline{e}LLLH}  - 2C^{{\tt (M),}wvrt}_{\overline{e}LLLH} + 4C^{{\tt (M),}wvtr}_{\overline{e}LLLH} \big)  + 2g^2 C^{prst}_{\overline{d}LQLH1} - \frac{2}{9} ( 10g^{\prime2} + 9g^2 - 24g^2_s ) C^{ptsr}_{\overline{d}LQLH1} 
	\nonumber
	\\
	& - C^{prvt}_{\overline{d}LQLH1} \big\{ 2 [ Y_{d} Y^\dagger_{d} ]_{vs}  - [ Y_{u} Y^\dagger_{u} ]_{vs}  \big\} + 2C^{pvst}_{\overline{d}LQLH1} [ Y_l Y^\dagger_l ]_{vr} - 3 [ Y^\dagger_{d} ]_{ps} [ Y_{d} ]_{vw} C^{wtvr}_{\overline{d}LQLH1}
	\nonumber
	\\
	&  - \frac{1}{36} ( 41 g^{\prime2} + 207g^2  + 96 g^2_s - 36W_H ) C^{prst}_{\overline{d}LQLH2} - \frac{2}{9} ( 10g^{\prime2} - 9g^2 - 24g^2_s ) C^{ptsr}_{\overline{d}LQLH2} 
	\nonumber
	\\
	& - \frac{1}{2} C^{prvt}_{\overline{d}LQLH2} \big\{ 3[ Y_{d} Y^\dagger_{d} ]_{vs} - 5[ Y_{u} Y^\dagger_{u} ]_{vs} \big\} + 3 C^{vrst}_{\overline{d}LQLH2} [ Y^\dagger_{d} Y_{d} ]_{pv} + \frac{5}{2} C^{prsv}_{\overline{d}LQLH2} [ Y_l Y^\dagger_l ]_{vt} 
	\nonumber
	\\
	& + \frac{1}{2} C^{pvst}_{\overline{d}LQLH2} [ Y_l Y^\dagger_l ]_{vr} + 3 [ Y^\dagger_{d} ]_{ps} [ Y_{d} ]_{vw} \big( C^{wrvt}_{\overline{d}LQLH2} - C^{wtvr}_{\overline{d}LQLH2} \big)  + 2C^{prvw}_{\overline{d}L u eH} [ Y^\dagger_{u} ]_{v s} [ Y^\dagger_l ]_{wt}
	\nonumber
	\\
	& + 2 \big\{ 3[ Y^\dagger_{d} ]_{ps} [ Y^\dagger_{u} ]_{wv} - [ Y^\dagger_{d} ]_{pv} [ Y^\dagger_{u} ]_{ws} \big\} C^{vwtr}_{\overline{Q}u L LH} 
    + \frac{1}{3} ( g^{\prime2} + 9g^2 ) C^{pvrt}_{\overline{d}uLDL} [ Y^\dagger_{u} ]_{vs} 
	\nonumber
	\\
	& + 2C^{pvrw}_{\overline{d}uLDL} [ Y^\dagger_{u} ]_{vs} [ Y_l Y^\dagger_l ]_{wt} \;,
	\\%%%
	\befun^{prst}_{\overline{d}L u e H} &= 2[ Y^\dagger_{d} Y_{u} ]_{ps}  \big\{ -3 C^{rt}_{L e DH} +  [ C_{DLDH1} Y_l ]^{rt} + 2 [ C_{DLDH2} Y_l ]^{rt} \big\}
	\nonumber
	\\
	&  + 2[ Y_{u} ]_{vs} [ Y_l ]_{wt} \big(- C^{prvw}_{\overline{d}LQLH1} +  C^{pwvr}_{\overline{d}LQLH1} + C^{prvw}_{\overline{d}L QLH2} + \frac{1}{2} C^{pwvr}_{\overline{d}L QLH2} \big) 
 \nonumber
	\\
	& - \frac{1}{4} (23 g^{\prime2} + 9 g^2 - 4W_H) C^{prst}_{\overline{d}L u e H} + 3 [ Y^\dagger_{d} Y_{d} ]_{pv} C^{vrst}_{\overline{d}L u e H}
    +  C^{pvsw}_{\overline{d}L u e H} [ Y_l ]_{vt} [ Y^\dagger_l ]_{wr}
	\nonumber
	\\
	& - \frac{3}{2} C^{pvst}_{\overline{d}L u e H} [ Y_l Y^\dagger_l ]_{vr}  
    + 3 C^{prvt}_{\overline{d}L u e H}  [ Y^\dagger_{u} Y_{u} ]_{v s}  
    + 2C^{prsv}_{\overline{d}L u e H} [ Y^\dagger_l Y_l ]_{vt}
	\nonumber
	\\
	&  + \big( C^{vsrw}_{\overline{Q}uLL H} + 2 C^{vswr}_{\overline{Q}uLL H} \big) [Y^\dagger_{d} ]_{pv} [ Y_l ]_{wt} - \frac{1}{3} ( 19 g^{\prime2} + 9g^2 ) C^{psrv}_{\overline{d}uLDL} [ Y_l ]_{vt} 
	\nonumber
	\\
	& - 2C^{psvw}_{\overline{d}uLDL} [ Y_l Y^\dagger_l ]_{vr} [ Y_l ]_{wt} + C^{vsrw}_{\overline{d}uLDL} [ Y^\dagger_{d} Y_{d} ]_{pv} [ Y_l ]_{wt}  +  C^{pvrw}_{\overline{d}uLDL} [ Y^\dagger_{u} Y_{u} ]_{vs} [ Y_l ]_{wt} \;,
	\\%%%
	\befun^{prst}_{\overline{Q}uLL H} &=  [Y_{u} ]_{pr} \big\{ 3g^2C^{st}_{DLDH1} + [ C_{DLDH1} Y_l Y^\dagger_l ]^{ts} + 2 [ C_{DLDH1} Y_l Y^\dagger_l ]^{st} \big\} 
	\nonumber
	\\
	& + \frac{1}{2}  [Y_{u} ]_{pr} \big\{ 3g^2C^{st}_{DLDH2} 
    +  4[ C_{DLDH2} Y_l Y^\dagger_l ]^{ts} - [ C_{DLDH2} Y_l Y^\dagger_l ]^{st} \big\} 
    \nonumber
	\\
	&-  [ Y_{u} ]_{pr} [ Y_l ]_{vw} \big( 3C^{{\tt (S),}wvst}_{\overline{e}LLLH} + 3C^{{\tt (A),}wvst}_{\overline{e}LLLH} 
    +  4C^{{\tt (M),}wvst}_{\overline{e}LLLH} - 2C^{{\tt (M),}wvts}_{\overline{e}LLLH} \big)
	\nonumber
	\\
	&  + \big( C^{vswt}_{\overline{d}L QLH 1} + C^{vswt}_{\overline{d}L QLH 2} - C^{vtws}_{\overline{d}L QLH 2} \big) \big\{ [ Y_{d} ]_{pv} [ Y_{u} ]_{wr} -  3 [ Y_{u} ]_{pr} [ Y_{d} ]_{wv}  \big\}
	\nonumber
	\\
	&  + C^{vtrw}_{\overline{d}L u e H} [ Y_{d} ]_{pv} [ Y^\dagger_l ]_{ws} + \frac{1}{12} ( g^{\prime2} -45 g^2 - 96 g^2_s + 12 W_H) C^{prst}_{\overline{Q}u LLH}
	\nonumber
	\\
	&  + 3 g^2 C^{prts}_{\overline{Q}u LLH}  + \frac{5}{2} C^{prvt}_{\overline{Q}u LLH}  [ Y_l Y^\dagger_l  ]_{vs} - \frac{1}{2} \big( 3C^{prsv}_{\overline{Q}u LLH} + 4 C^{prvs}_{\overline{Q}u LLH} \big) [ Y_l Y^\dagger_l  ]_{vt}  
	\nonumber
	\\
	& - \frac{1}{2} C^{vrst}_{\overline{Q}u LLH}  \big\{ [ Y_{d} Y^\dagger_{d} ]_{pv} - [ Y_{u} Y^\dagger_{u} ]_{pv} \big\} + C^{vrts}_{\overline{Q}u LLH} \big\{ 2 [ Y_{u} Y^\dagger_{u} ]_{pv} + [ Y_{d} Y^\dagger_{d} ]_{pv} \big\} 
	\nonumber
	\\
	& + 3 C^{pvst}_{\overline{Q}u LLH} [ Y^\dagger_{u} Y_{u} ]_{vr} + 6C^{wvst}_{\overline{Q}u LLH} [ Y^\dagger_{u} ]_{vw} [ Y_{u} ]_{pr}  + 3g^2 C^{vrst}_{\overline{d}u LDL} [ Y_{d} ]_{pv} 
	\nonumber
	\\
	& + 2C^{vrtw}_{\overline{d}u LDL} [ Y_{d} ]_{pv} [ Y_l Y^\dagger_l ]_{ws} \;,
	\\
	\befun^{prst}_{\overline{L} dud \widetilde{H}} &= - \frac{1}{12} ( 17 g^{\prime2} + 27 g^2 + 48 g^2_s - 12 W_H ) C^{prst}_{\overline{L} dud \widetilde{H}} - \frac{10}{3} g^{\prime2} C^{ptsr}_{\overline{L} dud \widetilde{H}} + 3 C^{pvst}_{\overline{L} dud \widetilde{H}} [ Y^\dagger_{d} Y_{d} ]_{vr} 
	\nonumber
	\\
	& + 3 C^{prsv}_{\overline{L} dud \widetilde{H}} [ Y^\dagger_{d} Y_{d} ]_{vt} - \frac{3}{2} C^{vrst}_{\overline{L} dud \widetilde{H}} [ Y_l Y^\dagger_l ]_{pv} + 2C^{prvt}_{\overline{L} dud \widetilde{H}} [ Y^\dagger_{u} Y_{u} ]_{vs} - 2\big( C^{{\tt (M),}prvt}_{\overline{L}dddH} 
	\nonumber
	\\
	& +  C^{{\tt (M),}pvrt}_{\overline{L}dddH} \big) [ Y^\dagger_{d} Y_{u} ]_{vs} + 4 C^{vwrt}_{\overline{e}Qdd\widetilde{H}} [ Y_l ]_{pv} [ Y_{u} ]_{ws} - 2\big( C^{prvw}_{\overline{L}dQQ\widetilde{H}} + C^{prwv}_{\overline{L}dQQ\widetilde{H}} \big) [ Y_{u} ]_{vs} [ Y_{d} ]_{wt} 
	\nonumber
	\\
	& + C^{vrtw}_{\overline{e}ddDd} [ Y^\dagger_{d} Y_{u} ]_{ws} [ Y_l ]_{pv} + \frac{1}{18} ( 29g^{\prime2} + 27g^2 + 96g^2_s ) C^{pvrt}_{\overline{L}QdDd} [ Y_{u} ]_{vs} 
	\nonumber
	\\
	& -  C^{pvrw}_{\overline{L}QdDd} \big\{  2[ Y^\dagger_{d} Y_{d} ]_{wt} [ Y_{u} ]_{vs} + [ Y^\dagger_{d} Y_{u} ]_{ws} [ Y_{d} ]_{vt}  \big\} + 2C^{pvwt}_{\overline{L}QdDd}  [ Y^\dagger_{d} Y_{d} ]_{wr} [ Y_{u} ]_{vs} \;,
	\\
	\befun^{{\tt (M),}prst}_{\overline{L}dddH} &= \frac{1}{6} [ Y^\dagger_{u} Y_{d} ]_{vr} \big( C^{ptvs}_{\overline{L}dud\widetilde{H}} -  C^{psvt}_{\overline{L}dud\widetilde{H}} \big) - \frac{1}{6} [Y^\dagger_{u} Y_{d} ]_{vs} \big( C^{ptvr}_{\overline{L}dud\widetilde{H}} + 2 C^{prvt}_{\overline{L}dud\widetilde{H}} \big) 
	\nonumber
	\\
	& + \frac{1}{6} [ Y^\dagger_{u} Y_{d} ]_{vt} \big( C^{psvr}_{\overline{L}dud\widetilde{H}} + 2  C^{prvs}_{\overline{L}dud\widetilde{H}} \big) - \frac{1}{12} ( 13g^{\prime2} + 27g^2 + 48 g^2_s - 12W_H ) C^{{\tt (M),}prst}_{\overline{L}dddH} 
	\nonumber
	\\
	& + 2C^{{\tt (M),}pvst}_{\overline{L}dddH} [ Y^\dagger_{d} Y_{d} ]_{vr} + 2C^{{\tt (M),}prvt}_{\overline{L}dddH} [ Y^\dagger_{d} Y_{d} ]_{vs} + 2C^{{\tt (M),}prsv}_{\overline{L}dddH} [Y^\dagger_{d} Y_{d} ]_{vt} 
	\nonumber
	\\
	& + \frac{5}{2} C^{{\tt (M),}vrst}_{\overline{L}dddH} [ Y_l Y^\dagger_l ]_{pv} + \frac{1}{2} [ Y_l ]_{pv} \big\{ C^{vwrs}_{\overline{e}ddDd} [ Y^\dagger_{d} Y_{d} ]_{wt} - C^{vwrt}_{\overline{e}ddDd} [ Y^\dagger_{d} Y_{d} ]_{ws}  \big\}
	\nonumber
	\\
	& - \frac{1}{9} ( 5g^{\prime2} - 12 g^2_s ) \big\{ C^{pvtr}_{\overline{L}QdDd} [ Y_{d} ]_{vs} - C^{pvsr}_{\overline{L}QdDd} [ Y_{d} ]_{vt} \big\} + \frac{1}{2} C^{pvwt}_{\overline{L}Q dDd} \big\{ [ Y_{d} ]_{vs} [ Y^\dagger_{d} Y_{d} ]_{wr} 
	\nonumber
	\\
	& +  [ Y_{d} ]_{vr} [ Y^\dagger_{d} Y_{d} ]_{ws} \big\} - \frac{1}{2} C^{pvws}_{\overline{L}Q dDd} \big\{ [ Y_{d} ]_{vt} [ Y^\dagger_{d} Y_{d} ]_{wr} + [ Y_{d} ]_{vr} [ Y^\dagger_{d} Y_{d} ]_{wt} \big\} 
	\nonumber
	\\
	& + \frac{1}{3} [ Y_l Y^\dagger_l ]_{pv} \big\{ C^{vwrs}_{\overline{L}Q dDd} [ Y_{d} ]_{wt} - C^{vwrt}_{\overline{L}Q dDd} [ Y_{d} ]_{ws}  \big\} \;,r
	\\%%%
	\befun^{prst}_{\overline{e}Qdd\widetilde{H}} &= \frac{1}{2} C^{vswt}_{\overline{L}dud\widetilde{H}} [ Y^\dagger_l ]_{pv} [ Y^\dagger_{u} ]_{wr} + \frac{1}{24} ( 11g^{\prime2} - 27g^2 - 48 g^2_s + 12 W_H ) C^{prst}_{\overline{e}Qdd\widetilde{H}} 
	\nonumber
	\\
	& +  C^{pvws}_{\overline{e}Qdd\widetilde{H}} [ Y_{d} ]_{vt} [ Y^\dagger_{d} ]_{wr} - \frac{1}{4} C^{pvst}_{\overline{e}Qdd\widetilde{H}} \big\{ 3[ Y_{u} Y^\dagger_{u} ]_{vr} - 5[ Y_{d} Y^\dagger_{d} ]_{vr} \big\} 
	\nonumber
	\\
	& + 3 C^{prvt}_{\overline{e}Qdd\widetilde{H}} [ Y^\dagger_{d} Y_{d} ]_{vs} + C^{vrst}_{\overline{e}Qdd\widetilde{H}} [ Y^\dagger_l Y_l ]_{pv} - \frac{1}{2} \big( C^{vsrw}_{\overline{L}dQQ\widetilde{H}} - 2 C^{vswr}_{\overline{L}dQQ\widetilde{H}}  \big) [ Y^\dagger_l ]_{pv} [ Y_{d} ]_{wt} 
	\nonumber
	\\
	& - g^{\prime2}  C^{pstv}_{\overline{e}ddDd} [ Y^\dagger_{d} ]_{vr} + \frac{3}{2} C^{pvsw}_{\overline{e}ddDd} [ Y^\dagger_{d} Y_{d} ]_{wt} [ Y^\dagger_{d} ]_{vr} - C^{vrsw}_{\overline{L}QdDd} [ Y^\dagger_{d} Y_{d} ]_{wt} [ Y^\dagger_l ]_{pv} 
	\nonumber
	\\
	& + C^{vwsx}_{\overline{L}QdDd} [ Y^\dagger_l ]_{pv} [ Y^\dagger_{d} ]_{xr} [ Y_{d} ]_{wt} - s \leftrightarrow t \;,
	\\
	\befun^{prst}_{\overline{L} dQQ\widetilde{H}} &= - \big(  2C^{prvw}_{\overline{L}dud\widetilde{H}} + C^{pwvr}_{\overline{L}dud\widetilde{H}} \big) \big\{ [ Y^\dagger_{u} ]_{v s} [ Y^\dagger_{d} ]_{wt} + [ Y^\dagger_{u} ]_{v t} [ Y^\dagger_{d} ]_{ws} \big\} - 2C^{vtwr}_{\overline{e}Qdd\widetilde{H}} [ Y_l ]_{pv} [ Y^\dagger_{d} ]_{ws} 
	\nonumber
	\\
	& - \frac{1}{12} ( 19g^{\prime2} + 45 g^2 + 48 g^2_s - 12W_H ) C^{prst}_{\overline{L}dQQ\widetilde{H}} - 3 g^2 C^{prts}_{\overline{L}dQQ\widetilde{H}} -  C^{pvwt}_{\overline{L}dQQ\widetilde{H}} [ Y^\dagger_{d} ]_{vs} [ Y_{d} ]_{wr} 
	\nonumber
	\\
	& - C^{pvsw}_{\overline{L}dQQ\widetilde{H}} [ Y^\dagger_{d} ]_{vt} [ Y_{d} ]_{wr} +  \frac{1}{2} \big( C^{vrst}_{\overline{L}dQQ\widetilde{H}} - 4 C^{vrts}_{\overline{L}dQQ\widetilde{H}} \big) [ Y_l Y^\dagger_l ]_{pv} +  3 C^{pvst}_{\overline{L}dQQ\widetilde{H}} [ Y^\dagger_{d} Y_{d} ]_{vr} 
	\nonumber
	\\
	& + \frac{1}{2} C^{prvt}_{\overline{L}dQQ\widetilde{H}} \big\{ [ Y_{d} Y^\dagger_{d} ]_{vs} + 5[ Y_{u} Y^\dagger_{u} ]_{vs}  \big\} + \frac{1}{2} \big( 4C^{prvs}_{\overline{L}dQQ\widetilde{H}} - 3C^{prsv}_{\overline{L}dQQ\widetilde{H}} \big) [ Y_{u} Y^\dagger_{u} ]_{vt} 
	\nonumber
	\\
	& + \frac{1}{2} \big( 5C^{prsv}_{\overline{L}dQQ\widetilde{H}} - 2C^{prvs}_{\overline{L}dQQ\widetilde{H}}  \big) [ Y_{d} Y^\dagger_{d} ]_{vt} + 3 C^{vwxr}_{\overline{e}ddDd} [ Y_l ]_{pv} [ Y^\dagger_{d} ]_{ws} [ Y^\dagger_{d} ]_{xt} 
	\nonumber
	\\
	& - \frac{2}{9} ( g^{\prime2} - 24 g^2_s ) C^{psrv}_{\overline{L}QdDd} [ Y^\dagger_{d} ]_{vt}  + 2C^{psvw}_{\overline{L}QdDd}  [ Y^\dagger_{d} Y_{d} ]_{vr} [ Y^\dagger_{d} ]_{wt} 
	\nonumber
	\\
	& -  C^{vtrw}_{\overline{L}QdDd} [ Y_l Y^\dagger_l ]_{pv} [ Y^\dagger_{d} ]_{ws} + C^{pvrw}_{\overline{L}QdDd}  \big\{ [ Y_{u} Y^\dagger_{u}]_{vs} [ Y^\dagger_{d} ]_{wt} + [ Y_{u} Y^\dagger_{u} ]_{vt} [ Y^\dagger_{d} ]_{ws} \big\} \;.
\end{align}
Note that we have corrected a typo in flavor indices of~\cite{Zhang:2023ndw} in the last three terms in \cref{eq:CeLLLHA} which involve $C_{\bar QuLLH}^{wvtr},~C_{\bar QuLLH}^{wvrs},~C_{\bar QuLLH}^{wvsr}$, so that both sides share the same flavor symmetry.
\\
\noindent$\bullet~\bm{\psi^4D}$
\begin{align}\label{eq:psi4D}
	\befun^{prst}_{\overline{d}uLDL} &= \big( 2 C^{st}_{DLDH1} + C^{st}_{DLDH2} \big) [ Y^\dagger_{d} Y_{u} ]_{pr} + \frac{1}{6} ( g^{\prime2} + 9g^2 ) C^{prst}_{\overline{d}uLDL} +  C^{vrst}_{\overline{d}uLDL} [ Y^\dagger_{d} Y_{d} ]_{pv} 
	\nonumber
	\\
	&+ C^{pvst}_{\overline{d}uLDL} [ Y^\dagger_{u} Y_{u} ]_{vr} + \frac{1}{2} C^{prvt}_{\overline{d}uLDL} [ Y_l Y^\dagger_l ]_{vs} + \frac{1}{2} C^{prsv}_{\overline{d}uLDL} [ Y_l Y^\dagger_l ]_{vt} \;,
	\\
	\befun^{prst}_{\overline{e}ddDd} &=  -\frac{2}{3} ( g^{\prime2} - 6g^2_s ) C^{prst}_{\overline{e}ddDd}  +  C^{vrst}_{\overline{e}ddDd}  [Y^\dagger_l Y_l ]_{pv} +  C^{pvst}_{\overline{e}ddDd} [ Y^\dagger_{d} Y_{d} ]_{vr} 
	\nonumber
	\\
	&+  C^{prvt}_{\overline{e}ddDd} [ Y^\dagger_{d} Y_{d} ]_{vs} + C^{prsv}_{\overline{e}ddDd} [ Y^\dagger_{d} Y_{d} ]_{vt}
	\nonumber
	\\
	& - \frac{2}{3} [ Y^\dagger_l ]_{pv}  \big\{ C^{vwst}_{\overline{L}QdDd} [ Y_{d} ]_{wr} + C^{vwrt}_{\overline{L}QdDd} [ Y_{d} ]_{ws} + C^{vwrs}_{\overline{L}QdDd} [ Y_{d} ]_{wt}  \big\} \;,
	\\
	\befun^{prst}_{\overline{L}QdDd} &= - 3 C^{vwst}_{\overline{e}ddDd} [ Y_l ]_{pv} [ Y^\dagger_{d} ]_{wr} + \frac{4}{9} \left( g^{\prime2} + 3g^2_s \right) C^{prst}_{\overline{L}QdDd} - C^{pvwt}_{\overline{L}QdDd} [ Y^\dagger_{d} ]_{wr} [ Y_{d} ]_{vs}
	\nonumber
	\\
	&  - C^{pvsw}_{\overline{L}QdDd} [ Y^\dagger_{d} ]_{wr} [ Y_{d} ]_{vt} + \frac{1}{2} C^{vrst}_{\overline{L}QdDd} [ Y_l Y^\dagger_l ]_{pv} + \frac{1}{2} C^{pvst}_{\overline{L}QdDd} \big\{ [ Y_{d} Y^\dagger_{d} ]_{vr} +  [ Y_{u} Y^\dagger_{u} ]_{vr} \big\}
	\nonumber
	\\
	& +  C^{prvt}_{\overline{L}QdDd}  [ Y_{d} Y^\dagger_{d} ]_{vs} +  C^{prsv}_{\overline{L}QdDd}  [ Y_{d} Y^\dagger_{d} ]_{vt} \;.
\end{align}

%%%%%%%%%%%%%%%%%%%%%%%%%%%%%%%%%%%%%%
\bibliography{references_paper.bib}{}
\bibliographystyle{JHEP}
%%%%%%%%%%%%%%%%%%%%%%%%%%%%%%%%%%%%%%		

\end{document}